\documentclass[a4paper,11pt]{article}

\usepackage{amsfonts,amssymb}
\usepackage{theorem}
\usepackage{epsf}
\usepackage{epsfig}
\usepackage{eufrak}
\begin{document}
\par
\rightline{IFUM-831-FT}
\vskip 0.4 truecm
\Large
\bf
\centerline{Endowing  the nonlinear sigma model  with a  flat}
\par
\centerline{ connection
structure: a way to renormalization}
\par
\centerline{ }
\par
\normalsize
\rm

\vskip 1.3 truecm
\large
\centerline{Ruggero Ferrari 
\footnote{E-mail address: {\tt ruggero.ferrari@mi.infn.it}} 
}

\vskip 0.3 truecm
\normalsize
\centerline{Physics Dept. University of Milan, } 
\centerline{via Celoria 16, 20133 Milan, Italy } 
\centerline{I.N.F.N., sezione di Milano}

\normalsize
\bf
\centerline{Abstract}

\rm
\begin{quotation}
We discuss the quantized theory of a pure-gauge non-abelian
vector field (flat connection) as it would appear in a
mass term {\sl \`a la} St\"uckelberg. However the paper
is limited to the case where only the flat connection is present
(no field strength term).
The perturbative solution is constructed by using only the
functional equations and by expanding in the number of
loops. In particular we do not use a perturbative approach
based on the path integral or on a canonical quantization.
It is shown that there is no solution with trivial S-matrix.
\par
Then the model is embedded in a nonlinear sigma model.
The solution is constructed by exploiting
a natural hierarchy in the functional
equations given by the number of insertions of the flat connection 
and of the constrained component of the sigma field. The amplitudes with
the sigma field are simply derived from those of the flat
connection and of the constraint component. 
Unitarity is enforced by hand by using Feynman
rules. We demonstrate the remarkable fact that in generic
dimensions the na{\"\i}ve Feynman rules yield amplitudes
that satisfy the functional equations. This allows a
dimensional renormalization of the theory in D=4 by recursive
subtractions of the poles in the Laurent expansion.
Thus one gets a finite theory depending only on two
parameters.
\par
The novelty of the paper is the use of the functional equation 
associated to the local left multiplication introduced by
Faddeev and Slavnov, here improved by adding the external
source coupled to the constrained component.
It gives a powerful tool to renormalize the nonlinear
sigma model.

\end{quotation}
\section{Introduction}
\label{sec:intr}
Gauge field theories have been very successful in the description
of the Physics of the sub-nuclear world. At the same
time remarkable progresses have been achieved on their
formulation, on the properties of the perturbative expansion,
on the relevance of the invariance principles and on
the importance of non-perturbative effects.  In this effort for
the  construction of a correct theoretical foundation 
a central r\^ole is played by the requirement of Physical Unitarity
\cite{unitarity}.
The story goes back to the Gupta-Bleuler formulation of
QED, where it is necessary to show that the unphysical
modes do decouple from the physical states. The same
strategy has been successful in the abelian gauge theory
in presence of spontaneous breakdown of the symmetry
(Higgs-Kibble model \cite{kibble}), where the ``gauge'' mode of vector
field conspires with the ``phase'' degree of freedom of
the scalar field in order to restore Physical Unitarity.
On the same road a solution has been found for the
problem of Physical Unitarity in non-abelian gauge theories
(also in presence of spontaneous breakdown). In this case
one had to wait for clever solution based on the
work of 't Hooft and Veltman\cite{'tHooft:1971rn}. This progress has been
possible after the introduction of the Faddeev-Popov ghosts
\cite{fp}.
The whole matter has been beautifully formulated after
the discovery of the properties of Becchi-Rouet-Stora-Tyutin 
\cite{BRST} symmetry. 
Also in this case the unphysical modes conspire in order to
save Physical Unitarity. 
\par
Unphysical modes are introduced into the gauge theories in the
process of quantization. Eventually they are described by
pure-gauge vector fields. 
\par
While there are no secrets about the properties of
the pure-gauge abelian field, the same cannot be said
for non-abelian gauge. If one introduces a mass term 
{\sl \`a la} St\"uckelberg \cite{stueckelberg}, a pure-gauge field (flat
connection) is needed that transforms according to 
\begin{eqnarray}
F'_{\mu}
= UF_{\mu}U^\dagger+ \frac{ i}{g}U\partial_\mu U^\dagger
\label{int.1}
\end{eqnarray}
and its  field strength is zero
\begin{eqnarray}
\partial_\mu F_\nu - \partial_\nu F_\mu -ig[F_\mu,F_\nu] =0.
\label{int.2}
\end{eqnarray}
The group transformations are generated from the algebra
\begin{eqnarray} 
[t_a,t_b] = i f_{abc}t_c,
\label{int.3}
\end{eqnarray}
\begin{eqnarray}
tr \left( t_a t_b \right) = \kappa\delta_{ab}
\label{int.4}
\end{eqnarray}
and
\begin{eqnarray}
F_\mu = t_a F_{a\mu}.
\label{int.5}
\end{eqnarray}
In this work we will study some general properties of the flat connection.
Then we will consider  its relation with the nonlinear sigma model
\cite{ketov, zinn-justin} and discuss the difficulties 
connected with the presence
of a non-trivial Haar measure 
and with the compact domain in the functional integration.
We suggest to abandon the usual perturbation theory based on 
gaussian integrals and to use instead  the functional
identities associated to the transformations in eq. (\ref{int.1}).
We exemplify our procedure at the tree level (which agrees
with a na{\"\i}ve approach). The one loop amplitudes are fixed
in order to guarantee unitarity of the S-matrix 
\cite{Cutkosky:1960sp,Veltman:1963th}.
Amplitudes up to four-point are evaluated. We discuss also
some two-loop amplitudes.
\par
The key discovery is that the na{\"\i}ve D-dimension Feynman
rules yield amplitudes that satisfy the functional
equation. This fact is demonstrated in few cases.
The importance of this fact is in the possibility of 
subtracting recursively the poles in the Laurent expansion,
thus getting a finite theory. The amplitudes are
given in terms of two parameters: coupling constant
and mass.
\par
The plan of the paper is the following. In Section \ref{sec:gen}
we briefly discuss the functional equations where only the flat connection
is present. We argue that the flat connection cannot describe
 a scalar particle with trivial dynamics ($S\not =1$).
In Section \ref{sec:sigma} the flat connection field is parametrized
by a nonlinear sigma model. The notations are fixed and the
Haar measure is discussed. In Section \ref{sec:str} we put in evidence
the difficulties in the quantization of the theory by using
perturbation theory in the path integral approach. Essentially
one cannot avoid the problems generated by the presence of the
Haar measure. In Section \ref{sec:del} we suggest a new strategy
from start: use the functional equations in order to define the
theory. In Section \ref{sec:renor} we conjecture that the na{\"\i}ve
Feynman rules yield a perturbative solutions of the functional
equations in generic D dimensions. This allows a recursive 
subtraction of the poles in $D=4$, thus yielding a finite theory.
This procedure of regularization is in the spirit of 
renormalization in the modern sense of \cite{Gomis:1995jp}.
In Section \ref{sec:hier} we discuss the hierarchy of the functional
equations, that can be ordered according to the number of
flat connections and other composite operators.
In Section \ref{sec:per} we discuss some standard aspects of
the loop-expansion and in  Section \ref{sec:tree} we evaluate
some tree-level amplitudes. In Section \ref{sec:rec} we recall the standard
procedure for the evaluation of connected amplitudes in terms of
those 1PI.
In Section \ref{sec:one} we
consider few one-loop  corrections and demonstrate
the essential point of the discovery, i.e. that the perturbative
expansion with standard Feynman rules in D dimensions yield
amplitudes that satisfy the functional equations. In Section \ref{sec:two}
we discuss the same problem for a two-loop two-point amplitude
and Section \ref{sec:more} we outline the proof for the general case.
In Section \ref{sec:D=4} we illustrate the subtraction procedure
in few examples.
\section{General properties}
\label{sec:gen}
Let us consider the path integral approach for the construction of
the functional
\begin{eqnarray} 
Z[J] = e^{iW[J]}=
\int {\cal D}[F]\exp i\int d^Dx\left (\frac{m_D^2}{8}F_{a\mu}F_a^\mu
+ F_{a\mu}J_a^\mu\right),
\label{gen.1}
\end{eqnarray}
where 
\begin{eqnarray} 
m_D \equiv m^\frac{D-2}{2}
\label{gen.1.0}
\end{eqnarray}
and
the functional measure is supposed to be invariant under
the transformations in eq. (\ref{int.1}). The complexity of the problem
is all hidden in the integration measure, since the action is a harmless
quadratic form in the field.
In general the integral cannot be 
performed analytically. Therefore we require only the validity of the
functional identity associated to the transformation in eq. (\ref{int.1}).
For the generating functional of the connected amplitude $W$ we get
\cite{FS}
\begin{eqnarray}
\Big(
\frac{1}{4}  m^2_D \partial^\mu  W_{ J_a^\mu}  
+ {\cal D}_{ab\mu}[ W_{ J}]J_b^\mu
\Big)(x)=0,
\label{gen.2}
\end{eqnarray}
where
\begin{eqnarray}&&
{\cal D}_{ab\mu}[X]\equiv \partial_\mu \delta_{ab} - g f_{abc}X_{c\mu}
\nonumber\\&&
{\cal D}_{\mu}[X]\omega \equiv
t_a {\cal D}_{ab\mu}[X]\omega_b = \partial_\mu \omega
- ig [X_{\mu},\omega].
\label{gen.3}
\end{eqnarray}
We use the simplified notation
\begin{eqnarray}&&
W_{a_1\dots a_n}^{\mu_1\dots\mu_n}(x_1,\dots,x_n)
= \frac{\delta^n W}{\delta J_{a_1\mu_1}(x_1)\dots \delta J_{a_n\mu_n}(x_n)}
\nonumber\\&&
\int d^Dx_1\dots d^Dx_n \exp i(p_1x_1+\dots+p_nx_n)
W_{a_1\dots a_n}^{\mu_1\dots\mu_n}(x_1,\dots,x_n)
\nonumber\\&&
=
W_{a_1\dots a_n}^{\mu_1\dots\mu_n}(p_1,\dots,p_n)
(2\pi)^D\delta(\sum_j p_j)
\label{gen.4}
\end{eqnarray}
The functional identity relates the connected amplitudes with different
number of points. The relation for $n>2$ is ($~\hat {}~$ means ``omitted'')
\begin{eqnarray}&&
\frac{m_D^2}{4} p_{k\mu_k}
W_{a_1\dots a_k \dots a_n}^{\mu_1\dots\mu_k\dots\mu_n}
(p_1,\dots, p_k,\dots,p_n)
\nonumber\\&&= i
g\sum_{j\not= k}f_{a_k a_j a'_j}
W_{a_1\dots a'_j\dots\hat a_k\dots a_n}
^{\mu_1\dots\mu_j\dots\hat\mu_k\dots\mu_n}
(p_1,\dots,p_j+p_k,\dots,\hat p_k,
\dots,p_n)
\label{gen.5}
\end{eqnarray}
For n=2 we have
\begin{eqnarray}
\frac{m_D^2}{4} p_{\mu}
W_{a b}^{\mu\nu}(p)
=-p^\nu \delta_{ab}.
\label{gen.6}
\end{eqnarray}
There is a trivial solution of eqs. (\ref{gen.5}) and (\ref{gen.6}), i.e.
all amplitudes are zero  except
\begin{eqnarray}
W_{a b}^{\mu\nu}(p)
= -\frac{4}{m^2_D} \delta_{ab}g^{\mu\nu}.
\label{gen.7}
\end{eqnarray}
This solution is formally given by the functional
\begin{eqnarray}
W[J] = - \int d^Dx \frac{2}{ m^2_D}J_a^\mu J_{a\mu}
\label{gen.8}
\end{eqnarray}
i.e. 
\begin{eqnarray}
Z[J] = \int {\cal D}[F] \exp i\int d^D x 
\left( \frac{m_D^2}{8} F_{a\mu}F_a^\mu 
+ J_{a\mu}F_a^\mu \right )
\label{gen.9}
\end{eqnarray}
where the integration is over the unconstrained configurations of the
classical field $F_{a\mu}$.
\par
The other possible solution is 
\begin{eqnarray}
W_{a b}^{\mu\nu}(p)
= \delta_{ab}\left((g^{\mu\nu}-\frac{p^\mu p^\nu}{p^2})W_T(p^2)-
\frac{4}{m^2_D} \frac{p^\mu p^\nu}{p^2}
\right)
\label{gen.10}
\end{eqnarray}
where the longitudinal part is exact. It would be tempting to construct
a solution where $W_T(p^2)=0$ and to recursively deduce 
all other amplitudes from eq. (\ref{gen.5}) with the constraint
that the S-matrix is equal one (no dynamics). 
\par
With these requirements one can easily construct the three point
function. One gets
\begin{eqnarray}&&
W_{a_1a_2a_3}^{\mu_1\mu_2\mu_3}(p_1,p_2,p_3) = -ig
\frac{8}{m^{4}_D}
f_{a_1a_2a_3}
\Big(\frac{p_1^{\mu_1}p_2^{\mu_2}}{p_1^2 p_2^2}(p_1-p_2)^{\mu_3}
\nonumber\\&&
+\frac{p_3^{\mu_3}p_2^{\mu_2}}{p_3^2 p_2^2}(p_2-p_3)^{\mu_1}
+\frac{p_1^{\mu_1}p_3^{\mu_3}}{p_1^2 p_3^2}(p_3-p_1)^{\mu_2}
\Big).
\label{gen.11}
\end{eqnarray}
However in evaluating the four point amplitude one finds
an obstruction: there is no solution of eq. (\ref{gen.2}) 
under the assumption that $W_T(p^2)=0$, i.e. trivial S-matrix.
Unfortunately the algebra
is rather cumbersome and it isn't worth to elaborate the explicit
calculation.
The next section provides better technical instruments in order
to investigate the whole problem of the construction of
the connected amplitudes.
\par
Since we cannot construct a trivial theory based on
eq. (\ref{gen.5}) we have to take into account radiative
corrections. Thus we consider a loop expansion where
the transverse part in eq. (\ref{gen.10}) is zero at the
tree level. With this assumption we evaluate the zero
loop amplitudes. Then we proceed to evaluate the one-loop
amplitudes by using the zero-loop results as Feynman rules
(unitarity). The best tool for these calculations is
provided by the 1PI amplitudes.


\section{Flat connection and nonlinear sigma model}
\label{sec:sigma}
The constraints in eqs. (\ref{int.1}) and (\ref{int.2}) can be
implemented by using a field $\Omega(x)$ with value in a 
unitary group $G$:
\begin{eqnarray}
F_\mu 
=\frac{ i}{g} \Omega\partial_\mu\Omega^\dagger=t_a F_{a\mu}
\label{sigma.1}
\end{eqnarray}
with
\begin{eqnarray}
\Omega^{-1}=\Omega^\dagger.
\label{sigma.2}
\end{eqnarray}
We parametrize the group with a set of real fields $\{\phi_a\}$.
The action in (\ref{gen.1}) becomes
\begin{eqnarray}&& 
S=
\int d^Dx \frac{m_D^2}{8}F_{a\mu}F_a^\mu
= \int d^Dx \frac{m_D^2}{8g^2\kappa}~tr ~
\left(\partial_\mu\Omega \partial^\mu\Omega^\dagger
\right)
\nonumber\\&&
= \int d^Dx \frac{m_D^2}{8g^2\kappa}~tr ~
\left(\frac{\partial\Omega}{\partial\phi_a} 
\frac{\partial\Omega^\dagger}{\partial\phi_b}
\right)\partial_\mu\phi_a \partial^\mu\phi_b
\nonumber\\&&
= \int d^Dx \frac{m_D^2}{2}~\eta_{ab}(\phi) ~
\partial_\mu\phi_a \partial^\mu\phi_b
\label{sigma.3}
\end{eqnarray}
where
\begin{eqnarray}
\eta_{ab} = \frac{1}{4g^2\kappa}~tr ~
\left(\frac{\partial\Omega}{\partial\phi_a} 
\frac{\partial\Omega^\dagger}{\partial\phi_b}
\right)
\label{sigma.4}
\end{eqnarray}
with the property (as a consequence of unitarity)
\begin{eqnarray} 
\eta_{ab} =\eta_{ba}.
\label{sigma.4.1}
\end{eqnarray}

The nonlinear sigma model is of particular relevance
in quantum field theory. It is the fundamental ingredient
of some phenomenological models. It appears as a component
or as a limit in theoretical models in field theory and
in statistical field theory.
\par

We consider the nonlinear sigma model for $SU(2)$ in order
to keep the notations as simple as possible. Most of the
results can be generalized to other groups. 
We introduce the parametrization
\begin{eqnarray}
\Omega = \frac{1}{m_D}\left[\phi_0 + ig\tau_a \phi_a\right] 
\qquad t_a = \frac{\tau_a}{2}
\qquad a=1,2,3.
\label{sigma.5}
\end{eqnarray}
Eq. (\ref{sigma.2}) and
\begin{eqnarray} 
\det\Omega=1
\label{sigma.5.0}
\end{eqnarray}
gives
\begin{eqnarray}
\phi_0^2+g^2\vec \phi^2
=m_D^2.
\label{sigma.6}
\end{eqnarray}
Then
\begin{eqnarray}
\eta_{ab} = m_D^{-2}\left(
g^2 \frac{\phi_a\phi_b}{\phi_0^2} + \delta_{ab}
\right),
\label{sigma.7}
\end{eqnarray}
\begin{eqnarray}&&
S = \frac{1}{2}\int d^D x \left ( 
\partial^\mu\phi_a\partial_\mu \phi_a+
g^2\frac{\phi_a\partial^\mu\phi_a
\phi_b\partial_\mu\phi_b}{\phi_0^2}
\right)
\nonumber\\&&
=\frac{1}{2}\int d^D x \left (
\partial^\mu\phi_a\partial_\mu \phi_a +
\frac{1}{g^2}
\partial^\mu\phi_0\partial_\mu\phi_0
\right)
\label{sigma.8}
\end{eqnarray}
and
\begin{eqnarray}
\eta\equiv \det \eta = \frac{1}{m_D^4\phi_0^2}.
\label{sigma.9}
\end{eqnarray}
\par
The constraint (\ref{sigma.6}) and the action (\ref{sigma.8})
are invariant under the global transformations
\begin{eqnarray}&&
\delta \phi_0 = - g^2 \frac{\delta\omega_a}{2}\phi_a
\nonumber\\&&
\delta \phi_a = \frac{\delta \omega_a}{2}\phi_0 +g \frac{\delta
\omega_c}{2}
\epsilon_{abc}\phi_b.
\label{sigma.10}
\end{eqnarray}
Both constraint and action are also invariant under
the global transformations
\begin{eqnarray}&&
\delta \phi_0 = 0
\nonumber\\&&
\delta \phi_a =  
\frac{\delta\alpha_c}{2}
\epsilon_{abc}\phi_b.
\label{sigma.11}
\end{eqnarray}
The transformations in eqs. (\ref{sigma.10}) and (\ref{sigma.11}) describe
the invariance of the model under $SU(2)\otimes SU(2)$ group of
transformations.
\par
The transformations in eq. (\ref{sigma.10}) are given by the
left multiplication
\begin{eqnarray}
\Omega' = U(\delta\omega)\Omega \simeq 
(1+i\frac{g}{2}\tau_a\delta\omega_a)\Omega.
\label{sigma.12}
\end{eqnarray}
The flat connection is
\begin{eqnarray}
F_{a\mu}
=\frac{ i}{g}tr(\tau_a \Omega\partial_\mu\Omega^\dagger)
=
\frac{2}{m_D^2}\left[(\phi_0\partial_\mu\phi_a-(\partial_\mu\phi_0)\phi_a)
+g\epsilon_{abc}(\partial_\mu\phi_b)\phi_c
\right].
\label{sigma.13}
\end{eqnarray}
The properties of the flat connection under local transformations 
(\ref{sigma.10}) is that of a gauge field (\ref{int.1})
\begin{eqnarray}
F'_{\mu}&&
= UF_{\mu}U^\dagger+ \frac{ i}{g}U\partial_\mu U^\dagger
\nonumber\\&&
\simeq F_{\mu} + ig [\omega, F_{\mu}] + \partial_\mu \omega, 
\qquad {\rm with }~ \omega = \frac{\tau_a}{2}\omega_a
\nonumber\\&&
= F_{\mu} + {\cal D}_\mu[F]\omega
\label{sigma.14}
\end{eqnarray}
where the covariant derivative is defined in (\ref{gen.3}).
\par
The transformations  given by the right multiplication
\begin{eqnarray}
\Omega' = \Omega U^\dagger(\delta\bar\omega)\simeq 
\Omega(1-i\frac{g}{2}\tau_a\delta\bar\omega_a)
\label{sigma.15}
\end{eqnarray}
yield
\begin{eqnarray}&&
\delta \phi_0 =  g^2\frac{\delta\bar\omega_a}{2}\phi_a
\nonumber\\&&
\delta \phi_a =  -\frac{\delta\bar\omega_a}{2}\phi_0 +g
\frac{\delta\bar\omega_c}{2}
\epsilon_{abc}\phi_b.
\label{sigma.16}
\end{eqnarray}
For ``{\sl local}'' right multiplication 
the flat connection transforms into new composite operators,
therefore they produce functional identities which are not useful
for the construction of the generating functionals.
\par
The paper concerns the construction of the n-point function
of the flat connection
\begin{eqnarray}
\langle 0|T(F_{a_1}^{\mu_1}(x_1)\cdots F_{a_n}^{\mu_n}(x_n))|0\rangle.
\label{sigma.17}
\end{eqnarray}
The  construction of  these n-functions will require
the study of the functions
\begin{eqnarray}
\langle 0|T(\phi_{a_1}(x_1)\cdots \phi_{a_n}(x_n))|0\rangle.
\label{sigma.18}
\end{eqnarray}
Thus we introduce
the external sources by the term
\begin{eqnarray}
S_{\rm sc}=\int d^D x (\phi_0 K_0 + \phi_a K_a + F_{a\mu} J_{a}^\mu).
\label{sigma.19}
\end{eqnarray}

In the paper we discuss the construction of the solution
of eq. (\ref{gen.2})
by considering the loop expansion of the connected amplitudes
and of the vertex functions (1PI amplitudes).
The iteration procedure is chosen in order to satisfy
unitarity of the scattering amplitude. The necessity of this procedure
is due to the complicated (non-polynomial) structure of the action
and on the fact that the path integral approach has a non-trivial
Haar measure in the functional integration.
The next sections are devoted to clarify the above points. 
%
%
%
\section{In straits}
\label{sec:str}
The importance of the nonlinear sigma model is hindered by
the difficulties present in the procedure of quantization.
The difficulties come mainly from the constraint on the field
components, under various aspects. Let us elaborate
on this point, since it is the starting point of our
approach.
The path integral formulation of the quantized theory
starts from a formal definition of the generating functional.
\par
The path integral approach to the quantization of the theory
needs the introduction of an invariant measure over the group
\begin{eqnarray} 
Z[K,K_0,J] = \int {\cal D}[\Omega]\exp
i(S+S_{\rm sc}).
\label{str.1}
\end{eqnarray}
In terms of parameter fields we have
\begin{eqnarray} 
Z[K,K_0,J] = \int \prod_x\sqrt{\det \eta(\phi)} d\phi(x)
\exp i(S+S_{\rm sc}).
\label{str.2}
\end{eqnarray}
For instance the measure in the case of equation
(\ref{sigma.9}) is given by
\begin{eqnarray}
 Z[K,K_0,J] = \prod_x \int_{g^2|\phi|^2<m_D^2}
\frac{d^3\phi(x)}{\phi_0}
\exp i(S+S_{\rm sc}).
\label{str.3}
\end{eqnarray}

The path integral in eq. (\ref{str.3}) shows some peculiarities:
a non-trivial Haar measure is present and the integration over the
field is on a compact region. These facts cause some
difficulties in formulating a perturbative expansion by using the 
path integral. We would like to illustrate these difficulties.
If we use a straightforward series expansion of the exponential
containing the "interaction", the quantities to be evaluated
are
\begin{eqnarray}
 \prod_x \int_{g^2|\phi|^2<m_D^2}
\frac{d^3\phi(x)}{\phi_0}\left(\phi_{a_1}(y_1)\dots\phi_{a_m}(y_m)
\right)
\exp(\frac{i}{2}\int d^Dz \partial_\mu\phi_a\partial^\mu\phi_a).
\label{str.3.1}
\end{eqnarray}
Here the integrals are not Gaussian and therefore the Wick 
expansion is in general not valid. The evaluation of the path
integrals over a group manyfold are in general beyond the present day 
technical ability, for anything beyond the two-point
function \cite{tome}.
\par
In alternative to the above expansion, it has been suggested that 
one should expand in the coupling
constant $g$. Therefore the Haar measure should be treated by the
exponentiation and moreover that the integration can be extended to
infinity.
A careful analysis of the divergences in the path integral formulation
has shown a remarkable cancellation between the divergent terms
coming from the Haar measure and some of the divergences of the action
\cite{early}. In particular the most severe divergent terms are
reabsorbed by a redefinition of the Weinberg's function 
$f(\vec\phi^2)$\cite{weinberg68}. This indicates that the principal r\^ole of
the action in the renormalization process has to be somehow supplemented
or enlarged to include a redefinition of the fields.
\par
A clear cut solution of the problem has been proposed after the discovery
of dimensional regularization \cite{dim}. In studying the renormalization
of the nonlinear sigma model in two dimension Brezin Zinn-Justin le Guillou
\cite{Brezin:1976ap}
discussed the problem of the non trivial Haar measure
and suggested that dimensional renormalization provides
a solution. On the same line 't Hooft \cite{'tHooft:1987rt} 
noticed that all the infinities
coming from the Haar measure could be disposed of by dimensional
renormalization. The approach based on the exponentiation of the
Haar measure deals with ill defined objects (usually $\delta(0)$)
and moreover the integration over the fields is taken on the
whole real axis by considering the approximation of small $g$ for the
limit of integration $\sim 1/g$. The final output of these
approximations is untenable, since the generating functional 
yields an equation of motion that is  not in agreement with the
exact functional equation derived from the local invariance of
the Haar measure, as discussed in the next Section (see in particular
the footnote on the eq. (\ref{del.2})).
\par
A further serious difficulty in the quantization of the nonlinear
sigma model was pointed out by T\u{a}taru \cite{Tataru:1975ys}. In the one-loop
amplitude in dimensional renormalization terms show up that violate
manifestly the global invariance. 
Appelquist and Bernard \cite{Appelquist:1980ae}
showed that these terms can be reabsorbed by reparametrization
of the scalar field. This reparametrization however includes
space-time derivatives i.e. is not of the form in eq. (\ref{sigma.3}).
\par
To the best of our knowledge this was the last contribution to
the problem of the quantization of the nonlinear sigma model.
However the model has been considered in many different
phenomenological schemes, where the problem of regularization
has been addressed. We mention the chiral field theory \cite{chiral}
where the Matching Conditions method \cite{espriu} has been
used.
In the next section we suggest  an alternative approach to
the construction of perturbative solution of the nonlinear
sigma model by using the n-point flat connection amplitudes.

\section{Deliverance}
\label{sec:del}
In the previous section we have pointed out the difficulty
in dealing with perturbation theory if a non-trivial Haar measure
is present in the path integral and the integration over the
fields is on a compact support. 
\par
We try to pursue a different approach. By starting from the
fact that our goal is the evaluation of the correlation functions,
we might try to get them by solving directly the functional equations
as an alternative to 
formal integration and subsequent  perturbative expansion 
in terms of Gaussian integrals.
\par
\begin{figure}
\epsfxsize=100mm
\centerline{\epsffile{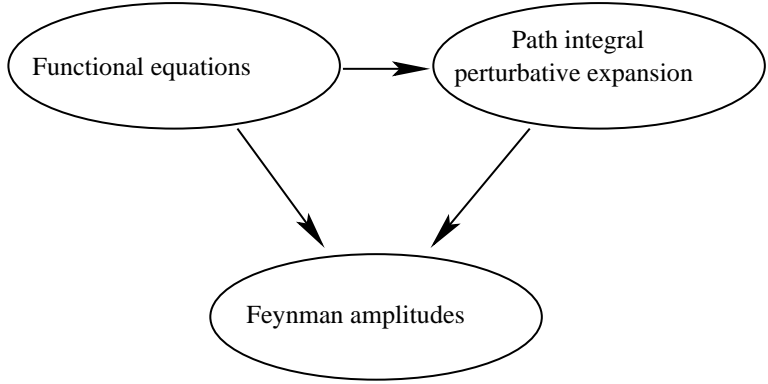}}
\caption{Alternatives in the construction of Feynman amplitudes}
\label{logical-graph}
\end{figure}
Since the functional given by functional integration in eq.
(\ref{str.1}) is a sum
over the group of transformations $\Omega$ we conclude that the
extremal should be found inside the set of configurations
$\{\Omega(x)\}$. Therefore the infinitesimal variation
of the field variables should be constrained to
the set given by eq. (\ref{sigma.6}). In general we consider
the variation (\ref{sigma.10})
\begin{eqnarray} 
\delta \Omega(x) = ig \delta\omega_a(x) t_a \Omega(x).
\label{del.1}
\end{eqnarray}
The identity for the generating functional $Z$ (or for that
of the connected $W$) is
\footnote{
Notice the difference with the usual equations of motion obtained
by a change of coordinates $\phi_a\to\phi_a+\delta\phi_a$
(for semplicity we take $J_{a\mu}=0$)
\begin{eqnarray}&&
\int {\cal D}[\vec \phi]\exp i(S+\int d^Dz K_0(z) \phi_0(z)) 
\left (
(\frac{1}{\phi_0}\Box\phi_0 \phi_a -\Box\phi_a)+K_a - g^2
\frac{1}{\phi_0}K_0 \phi_a
\right)(x)=0,
\nonumber\\
\label{str.4p }
\end{eqnarray}
which requires a set of new composite operator sources.
}
%
\begin{eqnarray}&& 
\int {\cal D}[\Omega]\exp(i( S+S_{\rm sc}))
\Big (
\frac{i }{4 g\kappa} tr \left ( -t_a \Omega\Box\Omega^\dagger
+ \Box \Omega\Omega^\dagger t_a 
\right)
\nonumber\\&&
- g^2\phi_a K_0 + \phi_0 K_a + g\epsilon_{abc}K_b\phi_c
+ 2{\cal D}_{ab\mu}[F]J_b^\mu
\Big)(x)
\nonumber\\&&
=\int {\cal D}[\Omega]\exp(i (S+S_{\rm sc}))
\Big (
 \left (\phi_a\Box \phi_0-  \Box \phi_a \phi_0
\right)  + g\epsilon_{abc}\phi_b\Box\phi_c
\nonumber\\&&
- g^2\phi_a K_0 + \phi_0 K_a + g\epsilon_{abc}K_b\phi_c
-2 {\cal D}_{ab\mu}[F]J_b^\mu
\Big)(x)=0 .
\label{del.2}
\end{eqnarray}
From eq. (\ref{sigma.13}) we get
\begin{eqnarray}&&
\int {\cal D}[\Omega]\exp(i( S+S_{\rm sc}))
\Big(-
\frac{ m_D^2}{2} \partial^\mu F_{a\mu} 
\nonumber\\&&
- g^2\phi_a K_0 + \phi_0 K_a+ g\epsilon_{abc}K_b\phi_c
- 2{\cal D}_{ab\mu}[F]J_b^\mu
\Big)(x)=0.
\label{del.3}
\end{eqnarray}
Thus the identity for the generating functional of the connected amplitudes
is (in $D$ dimensions)
\begin{eqnarray}&&
\Big(-
\frac{ m_D^2}{2}\partial^\mu \frac{ \delta W}{\delta J_a^\mu}  
- g^2  \frac{\delta W}{\delta K_a}  K_0 
+ \frac{\delta W}{\delta K_0} K_a
+ g\epsilon_{abc}K_b \frac{\delta W}{\delta K_c}
\nonumber\\&&
- 2{\cal D}_{ab\mu}[ \frac{\delta W}{\delta J}]J_b^\mu
\Big)(x)=0.
\label{del.4}
\end{eqnarray}
For the generating functional of the 1PI amplitudes one has
\begin{eqnarray}&&
\frac{\delta\Gamma[J,K_0,\vec\phi]}{\delta J_a^\mu}=
\frac{\delta W[J,K_0,\vec K]}{\delta J_a^\mu}
\nonumber\\&&
\frac{\delta\Gamma[J,K_0,\vec\phi]}{\delta K_0}=
\frac{\delta W[J,K_0,\vec K]}{\delta K_0}
\label{del.5}
\end{eqnarray}
and therefore
\begin{eqnarray}&&
\Big(
\frac{m_D^2}{2}\partial^\mu  \frac{ \delta \Gamma}{\delta J_a^\mu}
+g^2\phi_a K_0 +  \frac{ \delta \Gamma}{\delta K_0}
\frac{\delta\Gamma}{\delta \phi_a}
+g\epsilon_{abc} \frac{ \delta  \Gamma}{\delta \phi_b}\phi_c
\nonumber\\&&
+2{\cal D}[ \frac{ \delta \Gamma}{\delta J}]_{ab}^\mu J_{b\mu}\Big)(x)=0.
\label{del.6}
\end{eqnarray}
By using eqs. (\ref{del.4}) and (\ref{del.6}) we shall try to
find the n-point function of the flat connection   $F_\mu $.
Since we do not solve these equations with the help of
functional integration, unitarity has to be enforced at
every order in the loop expansion. This will be achieved by
performing {\sl contractions} on the external legs with
the Feynman prescription on the complex integration for
the propagators.
\par
Eqs. (\ref{del.4}) and (\ref{del.6}) establish an important
hierarchy in the construction of the perturbative solution.
Once the amplitudes for flat connections and for $\phi_0$ are evaluated,
those involving the field $\vec\phi$ are derived by using
eqs. (\ref{del.4}) and (\ref{del.6}) in a descending order:
at each step one gets one more $\vec\phi$ field and one less
flat connection. This somehow
takes into account automatically the arbitrariness in the parametrization
of the field $\Omega$ in terms of $\vec\phi$. Different
parameterizations yield different functional equations, i.e.
different Feynman rules involving the nonlinear sigma model.
\par
The renormalization via analytic continuation in the space-time
dimensions must take into account that eqs. (\ref{del.4}) and (\ref{del.6})
carry a dimensional parameter $m$. With a simple dimensional
argument one sees that the minimal subtraction (poles in $D=4$)
has to be done on the normalized amplitudes
\begin{eqnarray}
\left(
\frac{m_D}{m}
\right)^{2(n-1)}
\Gamma_{J_{a_1}^{\mu_1}\dots J_{a_n}^{\mu_n}}=
m^{(n-1)(D-4)}
\Gamma_{J_{a_1}^{\mu_1}\dots J_{a_n}^{\mu_n}}
.
\label{del.6.1}
\end{eqnarray}
For the amplitudes involving the fields $\vec\phi$ and $\phi_0$
eqs. (\ref{del.4}) and (\ref{del.6}) provide the correct $m_D$
factor.
\section{Renormalization}
\label{sec:renor}
In this section we present some personal view on renormalization.
We follow very close the idea of {\sl ``renormalization in the modern 
sense''} of Reference \cite{Gomis:1995jp}. 
We consider of particular interest the theories
that can be made finite in a symmetric fashion, i.e. by preserving the
salient properties as unitarity, Lorentz covariance, locality,
causality and in general  equations of motion and symmetries 
under the form of functional
identities for the generating functionals e.g. Ward identities,
Slavnov-Taylor identities, etc.. This is a pure mathematical
problem which might have many solutions. From the physical
point of view a particular solution can be proposed as a 
dynamical model on many different grounds, even that of elegance.
Any solution is in general a predictive model if it is supposed
to describe physical modes. However there are also other interesting
possibilities as for instance the case of theories where the
multiplicity of the solutions affects only the
unphysical modes.
\par
A subset of these  {\sl renormalizable} theories have
a unique perturbative solution once a finite number of 
parameters are fixed (i.e. those that
one indicates usually  as {``\sl renormalizable''}).
In the present work
we show that the nonlinear sigma model can be renormalized in
the wider meaning just introduced. 
\par
The amplitudes are expanded in the number of loops (or equivalently
in powers of the coupling constant $g$). The lowest terms of
this expansions are given by the tree graphs and can be obtained
directly by solving eq. (\ref{del.4}) or (\ref{del.6}). This
is exemplified in Section \ref{sec:tree}.
There is an alternative approach to the construction
of the tree level amplitudes, more close to our renormalization
strategy, as we outline in this section. By straightforward
calculation it can be proved that
\begin{eqnarray}&&
\Big(
\frac{m_D^2}{2}\partial^\mu  \frac{ \delta \Gamma^{(0)}}{\delta J_a^\mu}
+g^2\phi_a K_0 +  \frac{ \delta \Gamma^{(0)}}{\delta K_0}
\frac{\delta\Gamma^{(0)}}{\delta \phi_a}
+g\epsilon_{abc} \frac{ \delta  \Gamma^{(0)}}{\delta \phi_b}\phi_c
\nonumber\\&&
+2{\cal D}[ \frac{ \delta \Gamma^{(0)}}{\delta J}]_{ab}^\mu J_{b\mu}\Big)(x)=0,
\label{ren.1}
\end{eqnarray}
where the classical vertex functional is
\begin{eqnarray}
\Gamma^{(0)}
= \int d^D x \left \{ \frac{1}{2}\left (
\partial^\mu\phi_a\partial_\mu \phi_a +
\frac{1}{g^2}
\partial^\mu\phi_0\partial_\mu\phi_0
\right)
+K_0\phi_0 + J_a^\mu F_{a\mu}
\right\}.
\label{ren.2}
\end{eqnarray}
This result is very important since it saves a lot of work
in the explicit evaluation of the tree level amplitudes
for an arbitrary number of external legs. It is important
to notice that for $J_a^\mu =0$ the Feynman rules
derived from $\Gamma^{(0)}$ are symmetric under the transformation
\begin{eqnarray}
\phi_a \to - ~\phi_a.
\label{ren.2s}
\end{eqnarray}
\par
The higher loop amplitudes are constructed by using the
Feynman rules given by $\Gamma^{(0)}$ in eq.(\ref{ren.2}).
We shall demonstrate that this approach yields a solution
of eq. (\ref{del.6}) (or (\ref{del.4})), provided one neglects
all integrals representing a loop where no outside momentum flows
as shown in Refs. \cite{tadpole}. I.e.
\begin{eqnarray}
\int d^Dk \frac{1}{(k^2)^\alpha}=0,\qquad \alpha \in {\cal C}.
\label{ren.3}
\end{eqnarray}
We suggest that the amplitudes in $D=4$ dimensions are obtained
by iterative subtraction of the pole parts.
By subtracting the pole parts of the subgraphs according to the 
BPHZ prescription one obtains a finite amplitude for any n-1 loop 
amplitude. The pole part at one loop level is expected to satisfy 
the linearized functional equation
\begin{eqnarray}&&
\Big(
\frac{m_D^2}{2}\partial^\mu  \frac{ \delta }{\delta J_a^\mu}
+\frac{ \delta  \Gamma^{(0)}}{\delta \phi_a}
\frac{ \delta }{\delta K_0}
+ \frac{ \delta \Gamma^{(0)}}{\delta K_0}
\frac{ \delta }{\delta \phi_a}
+g\epsilon_{abc}\phi_c \frac{ \delta }{\delta \phi_b}
\nonumber\\&&
-2g\epsilon_{abc}J_{b\mu} \frac{ \delta }{\delta J_c^\mu}\Big)
\Gamma^{(1)}_{\rm POLE}=0.
\label{ren.4}
\end{eqnarray}
At higher order the counterterms $\hat\Gamma^{(n)}$ are fixed
according to hierarchy of eq. (\ref{del.6}).
\par
We summarize the renormalization strategy by the two-step procedure:
\begin{enumerate}
\item Enforce the validity of eqs. (\ref{del.4}) and (\ref{del.6})
in generic $D$ dimensions. This is achieved by neglecting all
tadpole contributions as required by the rule in eq. (\ref{ren.3}).
This point is crucial in our approach. Up to two loop the eq. (\ref{ren.3})
guarantees that the functional equations are satisfied
by the na{\"\i}ve Feynman rules. For higher number of loops
the situation is more complex and it is illustrated in Section \ref{sec:more},
where it is shown in a specific example that the equation of
motion given by the tree level of eq. (\ref{del.4})
\begin{eqnarray}&&
\Big(-
\frac{ m_D^2}{2}\partial^\mu \frac{ \delta W^{(0)}}{\delta J_a^\mu}  
- g^2  \frac{\delta W^{(0)}}{\delta K_a}  K_0 
+ \frac{\delta W^{(0)}}{\delta K_0} K_a
+ g\epsilon_{abc}K_b \frac{\delta W^{(0)}}{\delta K_c}
\nonumber\\&&
- 2{\cal D}_{ab\mu}[ \frac{\delta W^{(0)}}{\delta J}]J_b^\mu
\Big)(x)=0.
\label{del.4p}
\end{eqnarray}
restores the validity of eq. (\ref{del.6}) in D-dimensions.
\item Subtract recursively, as in BPHZ \cite{subtraction} renormalization 
procedure, the pole parts in the Laurent
expansion in $D-4$ by using the local solutions of eq. (\ref{ren.4})
at the one loop level. For higher order the subtraction procedure
requires a faithful acceptance of hierarchy. 
First fix the counterterms necessary for the subtraction of the poles in $D=4$
of the  amplitudes involving only flat connection
($F_a^\mu$) and $\phi_0$ (derivatives with respect to $K_0$) 
insertions normalized according to eq. (\ref{del.6.1}). 
Then derive the counterterms for amplitudes involving also 
$\vec\phi$.
\end{enumerate}
In Section \ref{sec:D=4}  we provide
some simple examples of this strategy.
It should be stressed once more that the key point is 
that the functional equations
(\ref{del.4}) and (\ref{del.6}) are valid in $D$
dimensions and therefore a symmetric subtraction is
possible. This statement is valid for theories that are
power counting renormalizable, where the constraint
on the dimension of the counterterms plays a
crucial r\^ole. Here we can only conjecture that the
removal of the poles at $D=4$ does not conflict with
the functional eqs.   (\ref{del.4}) and (\ref{del.6}).
In the present paper we verify the validity of the conjecture
on a two-loop example ($\Gamma_{JJ}^{(2)},
\Gamma_{J\phi}^{(2)}, \Gamma_{\phi\phi}^{(2)} $).
\par
We want to highlight once again few points about this strategy.
\begin{itemize}
\item 
Since $\Gamma^{(0)}$ is the 1PI functional then at the tree
level the maximum number of $K_0$ or $J_\mu^a$ insertions is
one as can be seen from eq. (\ref{ren.2}).
\item For any radiative corrections the pole subtraction
at the one loop level
corresponds to a local solution of eq. (\ref{ren.4}).
The same equation controls all the possible choices
of dimensional renormalization.
Therefore the study of the equation (\ref{ren.4})
is very important for the renormalization of
the non linear sigma model. By defining
\begin{eqnarray}&&
{\cal S }\equiv
\frac{m_D^2}{2}\partial^\mu  \frac{ \delta }{\delta J_a^\mu}
+\frac{ \delta  \Gamma^{(0)}}{\delta \phi_a}
\frac{ \delta }{\delta K_0}
+ \frac{ \delta \Gamma^{(0)}}{\delta K_0}
\frac{ \delta }{\delta \phi_a}
+g\epsilon_{abc}\phi_c \frac{ \delta }{\delta \phi_b}
\nonumber\\&&
-2g\epsilon_{abc}J_{b\mu} \frac{ \delta }{\delta J_c^\mu},
\label{ren.5}
\end{eqnarray}
different choices of the local counterterms  part $\widehat\Gamma$,
$\widehat\Gamma'$
have to satisfy the equation
\begin{eqnarray}
{\cal S}~\Bigl(\widehat\Gamma- \widehat\Gamma'\Bigr) =0.
\label{ren.6}
\end{eqnarray}
\item
After the subtractions have been performed the finite
amplitudes are given in terms of two parameters only:
$g$ and $m$. Thus in principle the theory becomes
predictive. However the dependence of the amplitudes from
the parameters $g$ and $m$ is entangled
with  the choice of the finite parts that satisfy eq. (\ref{ren.6}).
\end{itemize}

\par
\section{Hierarchy of the functional equations}
\label{sec:hier}
We will consider some n-point functions, that will be constructed
by using the functional equations. 
The solution of eqs. (\ref{del.4}) and (\ref{del.6})
is given in terms of some parameters. The invariance properties
of the equations allows to fix some of these at convenient values.
For instance the invariance under
\begin{eqnarray}&&
K_0 \to v K_0
\nonumber\\&&
\vec \phi \to \frac{1}{v} \vec \phi
\label{hier.1}
\end{eqnarray}
allows to fix the condition
\begin{eqnarray}
\left .
 W_{ K_0}\right |_{\vec K=K_0=J_\mu=0}=
\left .
\Gamma_{ K_0}\right |_{\vec K=K_0=J_\mu=0} =m_D.
\label{hier.2}
\end{eqnarray}
One can also consider the transformation
\begin{eqnarray}&&
J_{a\mu} \to g J_{a\mu} 
\nonumber\\&&
\vec \phi \to \frac{1}{g} \vec \phi
\nonumber\\&&
m_D \to g\,\,m_D
\label{hier.3}
\end{eqnarray}
which removes the constant $g$ in eq. (\ref{del.6}). However
this transformation induces a $g$-dependence in eq.
(\ref{hier.2}), which we prefer to avoid since eq. (\ref{hier.2})
fixes the spontaneous breakdown of the global symmetry under left
multiplication expressed by eq. (\ref{sigma.10}).
\par
We shall consider the explicit functional derivatives of eq.
(\ref{del.6}). This will show the hierarchy implicit in 
the functional equation: once the function for the flat-connection
and of the constrained component $\phi_0$
only is evaluated, the sigma model follows simply by successive
derivatives. We list some of them where use is made of the condition
(\ref{hier.2}).
\subsection{Two-point functions}
By taking one derivative of eq. (\ref{del.6})
\begin{eqnarray}
m_D
\partial^\mu \Gamma_{ J_a^\mu\phi_b}
+ 2\Gamma_{ \phi_a\phi_b}=0,
\label{hier.4}
\end{eqnarray}
and
\begin{eqnarray}
m_D^2
\partial^\mu \Gamma_{ J_a^\mu J_b^\nu}
+2m_D\,\,\Gamma_{ \phi_a J_b^\nu}
+ 4\partial_\nu \delta(x-y) \delta_{ab} =0.
\label{hier.5}
\end{eqnarray}
\subsection{Three-point functions}
By taking two derivatives we get
%
\begin{eqnarray}&&
\frac{m_D^2}{2}\partial^{\mu_1} \Gamma_
{ J_{a_1}^{\mu_1} \phi_{a_2} \phi_{a_3}}
+m_D\,\, \Gamma_
{ \phi_{a_1} \phi_{a_2} \phi_{a_3}}
\nonumber\\&&
+g\left(\epsilon_{{a_1}b{a_3}} 
\Gamma_{ \phi_b\phi_{a_2}}\delta(x_1-x_3)
-\epsilon_{{a_1}{a_2}b}\Gamma_{ \phi_b\phi_{a_3}}
\delta(x_1-x_2)
\right)
=0.
\label{hier.6}
\end{eqnarray}
%

%
\begin{eqnarray}&&
\frac{m_D^2}{2}\partial^{\mu_1} \Gamma_
{ J_{a_1}^{\mu_1} J_{a_2}^{\mu_2} \phi_{a_3}}
+ m_D\,\,\Gamma_
{ \phi_{a_1} J_{a_2}^{\mu_2} \phi_{a_3}}
\nonumber\\&&
+g\epsilon_{{a_1}b{a_3}} 
\Gamma_{ \phi_b J_{a_2}^{\mu_2}}\delta(x_1-x_3)
+2g\epsilon_{{a_1}b{a_2}} 
\Gamma_{ J_{b}^{\mu_2} \phi_{a_3}}\delta(x_1-x_2)
=0.
\label{hier.7}
\end{eqnarray}
%
%
\begin{eqnarray}&&
\frac{m_D^2}{2}\partial^{\mu_1} \Gamma_
{ J_{a_1}^{\mu_1} J_{a_2}^{\mu_2} J_{a_3}^{\mu_3}}
+ m_D\,\,\Gamma_
{ \phi_{a_1} J_{a_2}^{\mu_2} J_{a_3}^{\mu_3}}
\nonumber\\&&
+2g\epsilon_{{a_1}b{a_3}} 
\Gamma_{ J_{a_2}^{\mu_2} J_{b}^{\mu_3}}
\delta(x_1-x_3)
+2g\epsilon_{{a_1}b{a_2}} 
\Gamma_{ J_{b}^{\mu_2} J_{a_3}^{\mu_3}}
\delta(x_1-x_2)
=0.
\nonumber\\&&
\label{hier.8}
\end{eqnarray}
Similarly
%
%
\begin{eqnarray}&&
\frac{m_D^2}{2}\partial^{\mu_1} \Gamma_
{ J_{a_1}^{\mu_1} \phi_{a_2} K_0}
+g^2 \delta_{a_1a_2}\delta(x_1-x_2)\delta(x_1-x_3)
+\Gamma_{ K_0K_0}
\Gamma_{ \phi_{a_1} \phi_{a_2}}
\nonumber\\&&
+ m_D\,\,\Gamma_{ \phi_{a_1} \phi_{a_2} K_0}
=0
\label{hier.8.1}
\end{eqnarray}
%
and
\begin{eqnarray}&&
\frac{m_D^2}{2}\partial^{\mu_1} \Gamma_
{ J_{a_1}^{\mu_1} J_{a_2}^{\mu_2} K_0}
+m_D\Gamma_{ J_{a_1}^{\mu_1} \phi_{a_2} K_0}
+\Gamma_{ K_0K_0}
\Gamma_{ \phi_{a_1} J_{a_2}^{\mu_2}}
=0.
\label{hier.8.2}
\end{eqnarray}
\subsection{Four-point functions}
By taking three derivatives we get
%
\begin{eqnarray}&&
\frac{m_D^2}{2}\partial^{\mu_1}\Gamma_
{ J_{a_1}^{\mu_1} \phi_{a_2} \phi_{a_3}\phi_{a_4}}
+m_D\,\,\Gamma_
{ \phi_{a_1} \phi_{a_2}\phi_{a_3}\phi_{a_4}}
\nonumber\\&&
+\sum_{j=2}^4 
\Gamma_{K_0 \phi_{a_{j+1}}
 \phi_{a_{j+2}}}
\Gamma_
{ \phi_{a_1} \phi_{a_j}}
+\sum_{j=2}^4 g\epsilon_{a_1 b a_j}\Gamma_{\phi_b \phi_{a_{j+1}}
    \phi_{a_{j+2}}}
\delta(x_1-x_j)
=0.
\nonumber\\&&
\label{hier.9}
\end{eqnarray}
%
%
\begin{eqnarray}&&
\frac{m_D^2}{2}\partial^{\mu_1} \Gamma_
{ J_{a_1}^{\mu_1} J_{a_2}^{\mu_2} \phi_{a_3} \phi_{a_4}}
+m_D\,\, \Gamma_
{ \phi_{a_1} J_{a_2}^{\mu_2} \phi_{a_3} \phi_{a_4}}
\nonumber\\&&
+
\Gamma_{ K_0 \phi_{a_3}
 \phi_{a_4}}
\Gamma_
{ \phi_{a_1} J_{a_2}^{\mu_2}}
+\sum_{j=3,4}
\Big(
\Gamma_
{ K_0 J_{a_2}^{\mu_2} \phi_{a_j}}
\Gamma_
{ \phi_{a_1} \phi_{a_{(j+1)}}}
\nonumber\\&&
+g\epsilon_{{a_1}b{a_j}} 
\Gamma_{ \phi_b J_{a_2}^{\mu_2} \phi_{a_{(j+1)}}}\delta
(x_1-x_j)
\Big)
+2g\epsilon_{{a_1}b{a_2}} 
\Gamma_{ J_{b}^{\mu_2} \phi_{a_3} \phi_{a_4}}
\delta(x_1-x_2)
=0.
\nonumber\\&&
\label{hier.10}
\end{eqnarray}
%
%
\begin{eqnarray}&&
\frac{m_D^2}{2}\partial^{\mu_1} \Gamma_
{ J_{a_1}^{\mu_1} J_{a_2}^{\mu_2} J_{a_3}^{\mu_3} \phi_{a_4}}
+m_D\,\,\Gamma_
{ \phi_{a_1} J_{a_2}^{\mu_2} J_{a_3}^{\mu_3} \phi_{a_4}}
+
\Gamma_{ K_0 J_{a_2}^{\mu_2} \phi_{a_4}}\Gamma_{ \phi_{a_1} J_{a_3}^{\mu_3}}
\nonumber\\&&
+
\Gamma_{ K_0 J_{a_3}^{\mu_3} \phi_{a_4}}\Gamma_{ \phi_{a_1} J_{a_2}^{\mu_2}}
+\Gamma_{ K_0 J_{a_2}^{\mu_2} J_{a_3}^{\mu_3}}\Gamma_{ \phi_{a_1} \phi_{a_4}}
\nonumber\\&&
+g\epsilon_{{a_1}b{a_4}}\Gamma_{\phi_b J_{a_2}^{\mu_2} J_{a_3}^{\mu_3}}
\delta(x_1-x_4)
+2g\epsilon_{{a_1}b{a_2}} 
\Gamma_{ J_{b}^{\mu_2}J_{a_3}^{\mu_3} \phi_{a_4}}
\delta(x_1-x_2)
\nonumber\\&&
+2g\epsilon_{{a_1}b{a_3}} 
\Gamma_{ J_{b}^{\mu_3}J_{a_2}^{\mu_2} \phi_{a_4}}
\delta(x_1-x_3)
=0.
\label{hier.11}
\end{eqnarray}
Finally we take only derivatives respect to $J_{a\mu}$
%
%
\begin{eqnarray}&&
\frac{m_D^2}{2}\partial^{\mu_1} \Gamma_
{ J_{a_1}^{\mu_1} J_{a_2}^{\mu_2} J_{a_3}^{\mu_3} J_{a_4}^{\mu_4}}
+m_D\,\,\Gamma_
{ \phi_{a_1} J_{a_2}^{\mu_2} J_{a_3}^{\mu_3} J_{a_4}^{\mu_4}}
\nonumber\\&&
+\sum_{j=2}^4 \Gamma_
{ K_0 J_{a_{j+1}}^{\mu_{j+1}} J_{a_{j+2}}^{\mu_{j+2}}}
\Gamma_
{ \phi_{a_1} J_{a_{j}}^{\mu_{j}}}
+2g\sum_{j=2}^4\epsilon_{{a_1}b{a_j}} 
\Gamma_{ J_{b}^{\mu_j}
  J_{a_{j+1}}^{\mu_{j+1}}
  J_{a_{j+2}}^{\mu_{j+2}}}
\delta(x_1-x_j)
=0.
\nonumber\\&&
\label{hier.12}
\end{eqnarray}
%

\section{Perturbative solution}
\label{sec:per}

\par
In Section \ref{sec:gen} we have considered some properties of
the amplitudes for a flat connection. In particular we have 
discovered that the longitudinal part of the two-point function
gets no corrections from the higher terms in the loop expansion
\begin{eqnarray}W_
{ J_a^\mu
J_b^\nu}(p) =\delta_{ab}\left ((p^2 g_{\mu\nu}- p_\mu p_\nu)
W_{\rm T}(p^2)-
\frac{4}{ m_D^2}\frac{p_\mu p_\nu}{ p^2}
\right).
\label{per.5}
\end{eqnarray}
We expect  the transverse part to be non zero at higher loop corrections.
It is interesting to see how eq. (\ref{per.5}) can be valid even in a
theory with non-trivial dynamics. To see this we consider 1PI
amplitudes.  That is we use eqs. (\ref{del.4})
and (\ref{del.6}). For the generating functional of the connected
amplitudes we have
\begin{eqnarray}
-\frac{m_D^2}{2}\partial^\mu  W_{ J_a^\mu(x) K_b(y)}
+ \delta_{ab}\delta(x-y) W_{ K_0}=0
\label{per.7}
\end{eqnarray}
i.e.
\begin{eqnarray}
 W_{ J_a^\mu K_b}=
2 i\frac{p_\mu}{m_D p^2}\delta_{ab}.
\label{per.8}
\end{eqnarray}
For the 1PI amplitudes we get the following two-point functions
\begin{eqnarray}
m_D
\partial^\mu \Gamma_{ J_a^\mu\phi_b}
+ 2\Gamma_{ \phi_a\phi_b}=0,
\label{per.10}
\end{eqnarray}
and
\begin{eqnarray}
m_D^2
\partial^\mu \Gamma_{ J_a^\mu J_b^\nu}
+2m_D\,\,\Gamma_{ \phi_a J_b^\nu}
+ 4\partial_\nu \delta(x-y) \delta_{ab} =0.
\label{per.11}
\end{eqnarray}
We can now reconstruct the connected amplitude in terms of the 1PI
ones.

\begin{eqnarray}
 W[J,K_0,\vec K]_{ J_a^\mu J_b^\nu}
=
\Gamma[J,K_0,\vec\phi]_{ J_a^\mu J_b^\nu}
+ W[J,K_0,\vec K]_{ J_a^\mu K_c}
\Gamma[J,K_0,\vec\phi]_{\phi_c  J_b^\nu}
\label{per.12}
\end{eqnarray}
The longitudinal part satisfies the equation (\ref{gen.6}):
\begin{eqnarray}&& \partial^\mu
 W[J,K_0,\vec K]_{ J_a^\mu J_b^\nu}
=\partial^\mu
\Gamma[J,K_0,\vec\phi]_{ J_a^\mu J_b^\nu}
+\partial^\mu W[J,K_0,\vec K]_{ J_a^\mu K_c}
\Gamma[J,K_0,\vec\phi]_{\phi_c  J_b^\nu}
\nonumber\\&&
=
-\frac{2}{m_D} \Gamma_{ \phi_a J_b^\nu}
- \frac{4}{m_D^2}\partial_\nu \delta(x-y) \delta_{ab}
+\int \frac{2d^D z}{m_D}\delta(x-z) \delta_{ac}
\Gamma_{ \phi_c(z) J_b^\nu}
\nonumber\\&&
=
-\frac{4}{m_D^2}\partial_\nu \delta(x-y) \delta_{ab}
\label{per.13}
\end{eqnarray}
where we have used eqs. (\ref{per.7}) and
(\ref{hier.5}). The above equation shows the cancellation of
the higher loop contributions. 

\subsection{Feynman rules}

As discussed in Sec. \ref{sec:hier} we first construct the
amplitudes where only the flat connection and the constrained 
component $\phi_0$ are involved and
then we use the functional eqs. (\ref{del.4}) and (\ref{del.6})
in order to obtain all the other (where also the $\phi_a$ appears).
In the first step we borrow the Feynman rules of the perturbative
path integral in order to exploit the properties that are
automatically provided by this scheme as unitarity, correct symmetry
factors, etc. That means:
\begin{itemize}
\item Vertexes:
\begin{eqnarray}
i\, \Gamma^{(0)}_{J^\mu_a,\dots,K_0,\dots,\phi_a,\dots}
\label{per.14}
\end{eqnarray}
\item Propagator:
\begin{eqnarray}
-i\, W^{(0)}_{K_aK_b}
= \frac{i}{p^2+i\epsilon}\delta_{ab}
\label{per.15}
\end{eqnarray}
\item Integration on internal lines:
\begin{eqnarray}
\frac{1}{(2\pi)^D}\int d^D k
\label{per.16}
\end{eqnarray}
\item Symmetry factors.

\end{itemize}

\section{Solution at the tree level}
\label{sec:tree}
In this section we construct some amplitudes at the tree level.
This is a redundant labor since we know already a solution
of eq. (\ref{del.6}) i.e. the classical action (\ref{ren.2}).
Functional derivatives of the classical action give all possible
vertexes. Here 
the aim is to show how the eqs. (\ref{del.4}) and (\ref{del.6})
can fix completely the solution. We require that at the tree level
\begin{eqnarray}
\left .
\frac{\delta^n\Gamma^{(0)}[J,K_0,\vec\phi]}
{\delta J_{a}^{\mu}\dots\delta  K_0\dots}\right|_{J=K_0=\vec K=0}
 = 0,
\label{tree.1}
\end{eqnarray}
whenever there is more than one insertion of composite operators
($F_\mu$ or $\phi_0$). With this {\sl Ansatz} we try to determine
the solution at the tree level. 
We will find up to four point amplitude that the solution agrees 
with a na{\"\i}ve reading of the action in eq. (\ref{ren.2}).
\subsection{Two-point functions}
From eqs. (\ref{hier.4}), (\ref{hier.5}) and (\ref{tree.1})
we get
\begin{eqnarray}
\Gamma^{(0)}_{ J_a^\mu\phi_b}= -
i2 \frac{ p_\mu}{ m_D}\delta_{ab}
\label{tree.2}
\end{eqnarray}
and
\begin{eqnarray}
\Gamma^{(0)}_{ \phi_a \phi_b} = 
\delta_{ab}p^2.
\label{tree.3}
\end{eqnarray}
\subsection{Three-point functions}

Take now further derivatives of eq. (\ref{del.6})
%
\begin{eqnarray}&&
\frac{m_D^2}{2}\partial^{\mu_1} \Gamma^{(0)}_
{ J_{a_1}^{\mu_1} \phi_{a_2} \phi_{a_3}}
+ m_D\,\,\Gamma^{(0)}_
{ \phi_{a_1} \phi_{a_2} \phi_{a_3}}
\nonumber\\&&
+g\left(\epsilon_{{a_1}b{a_3}} 
\Gamma^{(0)}_{ \phi_b\phi_{a_2}}\delta(x_1-x_3)
-\epsilon_{{a_1}{a_2}b}\Gamma^{(0)}_{ \phi_b\phi_{a_3}}
\delta(x_1-x_2)
\right)
=0.
\label{tree.3.1}
\end{eqnarray}
%

%
\begin{eqnarray}&&
\frac{m_D^2}{2}\partial^{\mu_1} \Gamma^{(0)}_
{ J_{a_1}^{\mu_1} J_{a_2}^{\mu_2} \phi_{a_3}}
+ m_D\,\,\Gamma^{(0)}_
{ \phi_{a_1} J_{a_2}^{\mu_2} \phi_{a_3}}
\nonumber\\&&
+g\epsilon_{{a_1}b{a_3}} 
\Gamma^{(0)}_{ \phi_b J_{a_2}^{\mu_2}}\delta(x_1-x_3)
+2g\epsilon_{{a_1}b{a_2}} 
\Gamma^{(0)}_{ J_{b}^{\mu_2} \phi_{a_3}}\delta(x_1-x_2)
=0.
\label{tree.4}
\end{eqnarray}
%
%
\begin{eqnarray}&&
\frac{m_D^2}{2}\partial^{\mu_1} \Gamma^{(0)}_
{ J_{a_1}^{\mu_1} J_{a_2}^{\mu_2} J_{a_3}^{\mu_3}}
+ m_D\,\,\Gamma^{(0)}_
{ \phi_{a_1} J_{a_2}^{\mu_2} J_{a_3}^{\mu_3}}
\nonumber\\&&
+2g\epsilon_{{a_1}b{a_3}} 
\Gamma^{(0)}_{ J_{a_2}^{\mu_2} J_{b}^{\mu_3}}
\delta(x_1-x_3)
+2g\epsilon_{{a_1}b{a_2}} 
\Gamma^{(0)}_{ J_{b}^{\mu_2} J_{a_3}^{\mu_3}}
\delta(x_1-x_2)
=0.
\nonumber\\&&
\label{tree.5}
\end{eqnarray}
From eqs. (\ref{tree.1}) and  (\ref{hier.7}) one gets
\begin{eqnarray}
\Gamma^{(0)}_
{ J_{a_1}^{\mu_1} \phi_{a_2} \phi_{a_3}}=
 2 i\frac{g}{m_D^2}
\epsilon_{{a_1}{a_2}{a_3}} (p_{2\mu_1}-p_{3\mu_1})
\label{tree.6}
\end{eqnarray}
and from eq.  (\ref{tree.3})

\begin{eqnarray}
\Gamma^{(0)}_
{ \phi_{a_1} \phi_{a_2} \phi_{a_3}} =0.
\label{tree.7}
\end{eqnarray}

By means of the 1PI amplitudes (\ref{tree.2}), (\ref{tree.3})
and (\ref{tree.6}) one can easily construct the connected amplitude
given in eq. (\ref{gen.11}).
\par
Finally by differentiating eq. (\ref{del.6}) with respect to $K_0,\phi_b$
we get
\begin{eqnarray}
\frac{m_D^2}{2}\partial^{\mu} \Gamma^{(0)}_
{ J_{a}^{\mu} K_0 \phi_b}
+
  m_D\,\,\Gamma^{(0)}_
{ K_0  \phi_{a} \phi_{b}} 
+g^2 \delta_{ab}=0.
\label{tree.7.1}
\end{eqnarray}
and similarly
\begin{eqnarray}
\frac{m_D^2}{2}\partial^{\mu_1} \Gamma^{(0)}_
{ J_{a_1}^{\mu_1} J_{a_2}^{\mu_2} K_0}
+
\Gamma^{(0)}_
{ K_0 K_0 }
 \Gamma^{(0)}_
{ \phi_{a_1} J_{a_2}^{\mu_2}} 
+\Gamma^{(0)}_
{ K_0}\Gamma^{(0)}_
{ \phi_{a_1}  J_{a_2}^{\mu_2} K_0}=0.
\label{tree.7.2}
\end{eqnarray}
At the tree level we get
\begin{eqnarray}
\Gamma^{(0)}_
{ K_0  \phi_{a} \phi_{b}} =
-\frac{g^2}{ m_D}\delta_{ab}.
\label{tree.8}
\end{eqnarray}
\subsection{Four-point functions}

Now we consider four-point functions. We take all possible derivatives
of eq. (\ref{del.6}). We use some simplification in the notations
%
\begin{eqnarray}&&
\frac{m_D^2}{2}\partial^{\mu_1}\Gamma^{(0)}_
{ J_{a_1}^{\mu_1} \phi_{a_2} \phi_{a_3}\phi_{a_4}}
+  m_D\,\,\Gamma^{(0)}_
{ \phi_{a_1} \phi_{a_2}\phi_{a_3}\phi_{a_4}}
\nonumber\\&&
+\sum_{j=2}^4 
\Gamma^{(0)}_{K_0 \phi_{a_{j+1}}
 \phi_{a_{j+2}}}
\Gamma^{(0)}_
{ \phi_{a_1} \phi_{a_j}}
+\sum_{j=2}^4 g\epsilon_{a_1 b a_j}\Gamma^{(0)}_{\phi_b \phi_{a_{j+1}}
    \phi_{a_{j+2}}}
\delta(x_1-x_j)
=0.
\nonumber\\&&
\label{tree.9}
\end{eqnarray}
%

In the above equation at the tree level the last
term should be zero according to eq. (\ref{tree.7}).
\begin{eqnarray}&&
\frac{m_D^2}{2}\partial^{\mu_1} \Gamma^{(0)}_
{ J_{a_1}^{\mu_1} J_{a_2}^{\mu_2} \phi_{a_3} \phi_{a_4}}
+  m_D\,\,\Gamma^{(0)}_
{ \phi_{a_1} J_{a_2}^{\mu_2} \phi_{a_3} \phi_{a_4}}
\nonumber\\&&
+
\Gamma^{(0)}_{ K_0 \phi_{a_3}
 \phi_{a_4}}
\Gamma^{(0)}_
{ \phi_{a_1} J_{a_2}^{\mu_2}}
+\sum_{j=3,4}
\Big(
\Gamma^{(0)}_
{ K_0 J_{a_2}^{\mu_2} \phi_{a_j}}
\Gamma^{(0)}_
{ \phi_{a_1} \phi_{a_{j+1}}}
\nonumber\\&&
+g\epsilon_{{a_1}b{a_j}} 
\Gamma^{(0)}_{ \phi_b J_{a_2}^{\mu_2} \phi_{a_{j+1}}}\delta
(x_1-x_j)
\Big)
+2g\epsilon_{{a_1}b{a_2}} 
\Gamma^{(0)}_{ J_{b}^{\mu_2} \phi_{a_3} \phi_{a_4}}
\delta(x_1-x_2)
=0.
\nonumber\\&&
\label{tree.10}
\end{eqnarray}
According to the {\sl Ansatz} in eq. (\ref{tree.1}) the first term
in eq. (\ref{hier.10}) is zero at the tree level.
%
\begin{eqnarray}&&
\frac{m_D^2}{2}\partial^{\mu_1} \Gamma^{(0)}_
{ J_{a_1}^{\mu_1} J_{a_2}^{\mu_2} J_{a_3}^{\mu_3} \phi_{a_4}}
+ m_D\,\,\Gamma^{(0)}_
{ \phi_{a_1} J_{a_2}^{\mu_2} J_{a_3}^{\mu_3} \phi_{a_4}}
\nonumber\\&&
+
\Gamma^{(0)}_
{ K_0 J_{a_2}^{\mu_2} \phi_{a_4}}
\Gamma^{(0)}_
{ \phi_{a_1} J_{a_3}^{\mu_3}}+
\Gamma^{(0)}_
{ K_0 J_{a_3}^{\mu_3} \phi_{a_4}}
\Gamma^{(0)}_
{ \phi_{a_1} J_{a_2}^{\mu_2}}
\nonumber\\&&
+\Gamma^{(0)}_
{ K_0 J_{a_2}^{\mu_2} J_{a_3}^{\mu_3}}
\Gamma^{(0)}_
{ \phi_{a_1} \phi_{a_4}}
=0.
\label{tree.11}
\end{eqnarray}
Finally we take only derivatives respect to $J_{a\mu}$
%
%
\begin{eqnarray}&&
\frac{m_D^2}{2}\partial^{\mu_1} \Gamma^{(0)}_
{ J_{a_1}^{\mu_1} J_{a_2}^{\mu_2} J_{a_3}^{\mu_3} J_{a_4}^{\mu_4}}
+ m_D\,\,\Gamma^{(0)}_
{ \phi_{a_1} J_{a_2}^{\mu_2} J_{a_3}^{\mu_3} J_{a_4}^{\mu_4}}
\nonumber\\&&
+\sum_{j=2}^4 \Gamma^{(0)}_
{ K_0 J_{a_{j+1}}^{\mu_{j+1}} J_{a_{j+2}}^{\mu_{j+2}}}
\Gamma^{(0)}_
{ \phi_{a_1} J_{a_{j}}^{\mu_{j}}}
+2g\sum_{j=2}^4\epsilon_{{a_1}b{a_j}} 
\Gamma^{(0)}_{ J_{b}^{\mu_j}
  J_{a_{j+1}}^{\mu_{j+1}}
  J_{a_{j+2}}^{\mu_{j+2}}}
\delta(x_1-x_j)
=0.
\nonumber\\&&
\label{tree.12}
\end{eqnarray}
From eq. (\ref{tree.10}) we get

\begin{eqnarray}&&
\Gamma^{(0)}_
{ J_{a_1}^{\mu_1} \phi_{a_2} \phi_{a_3} \phi_{a_4}}
=- i2\frac{g^2}{m^3_D}\Big( 
(p_1+2p_2)_{\mu_1}\delta_{a_3a_4}\delta_{a_1a_2}
\nonumber\\&& +
(p_1+2p_3)_{\mu_1}\delta_{a_2a_4}\delta_{a_1a_3} +
(p_1+2p_4)_{\mu_1}\delta_{a_2a_3}\delta_{a_1a_4}
\Big).
\nonumber\\&&
\label{tree.13}
\end{eqnarray}
and from eq. (\ref{tree.9})

\begin{eqnarray}&&
\Gamma^{(0)}_
{ \phi_{a_1} \phi_{a_2} \phi_{a_3} \phi_{a_4}}
=\frac{g^2}{m_D^2}
\Big( \delta_{a_3a_4}\delta_{a_1a_2}( p_2+p_1)^2
\nonumber\\&& +
\delta_{a_2a_4}\delta_{a_1a_3}( p_3+p_1)^2 +
\delta_{a_2a_3}\delta_{a_1a_4}( p_4+p_1)^2
\Big)
.
\nonumber\\&&
\label{tree.14}
\end{eqnarray}
Eq. (\ref{tree.14}) says that the flat connection indeed describes
a scalar particle which interacts with itself.
\par
From the brief analysis performed up to now one sees
that there is a natural hierarchy in the equations. From
the functions with the highest number of fields of the
flat connection and the constrained 
component $\phi_0$ one derives those where the scalar field
is involved.
%
%
%
\section{Reconstruction of the connected amplitudes}
\label{sec:rec}
The connected amplitudes can be constructed in
a straightforward way by using the obtained results
for the 1PI amplitudes. The starting point is given
by the relations in eqs. (\ref{del.5}). By using
the chain rules for the functional derivatives respect
to $J_{a\mu}$ with fixed $K_a$
\begin{eqnarray}&&
\frac{\delta\phi_z}{\delta J_1}=
 W_{ J_1 K_z}=
 \Gamma_{ J_1 \phi_w}
 W_{ K_w K_z}
= - \Gamma_{ J_1 \phi_w}
\left[
 \Gamma_{ \phi_w \phi_z}
\right]^{-1}
\nonumber\\&&
\label{rec.1}
\end{eqnarray}
and (see eq. (\ref{per.12}))
\begin{eqnarray}&&
W_{ J_a^\mu J_b^\nu}
=
\Gamma_{ J_a^\mu J_b^\nu}
+\Gamma_{ J_a^\mu \phi_c}
\frac{\delta\phi_c}{\delta J_b^\nu}
\nonumber\\&&
=
\Gamma_{ J_a^\mu J_b^\nu}
+\Gamma_{ J_a^\mu \phi_c}
 W_{ K_c K_w}
 \Gamma_{ \phi_w   J_b^\nu}.
\label{rec.2}
\end{eqnarray}
A further derivative respect to $J_{a\mu}$ with fixed $K_a$
is straightforward. The only new quantity is
\begin{eqnarray}&&
\frac{\delta
}{\delta J_a^\mu}
 W_{ K_c K_w}
=
 \Gamma_{ J_a^\mu K_c K_w}
=  \Gamma_{ J_a^\mu \phi_{c'}\phi_{w'}}
 W_{ K_{c'} K_c}
 W_{ K_{w'} K_w}.
\label{rec.3}
\end{eqnarray}
Thus we have simple rules in order to construct the full connected
amplitude with $n+1$ external legs of the flat connection:
\begin{enumerate}
\item{} Start form the n-derivative of $W$ where all the amplitudes 
are expressed in terms of 1PI functions, with the exception of the
$\phi-\phi$ propagator for which it is convenient to use the connected
amplitude.
\item{} Perform the derivative respect to $J_{a\mu}$ with fixed
  $K_a$ and put subsequently all the external sources to zero.
To this purpose the 1PI amplitudes are necessary, 
together with the identities
in eqs. (\ref{rec.1})) and (\ref{rec.3})).
\item{} For the final expression, it is convenient to perform
the replacement (see eq. (\ref{rec.1}))
\begin{eqnarray}
 \Gamma_{ J_1 \phi_w}
 W_{ K_w K_z}
\to
 W_{ J_1 K_z}.
\label{rec.4}
\end{eqnarray}
\end{enumerate}

\section{One-loop}
\label{sec:one}
The radiative corrections to the tree level
amplitudes is the main problem to be solved.
There are few items to be discussed. 
First of all we need a recipe for the construction of the flat connection
amplitudes, i.e. those that are at the top of the hierarchy.
We use the Feynman rules introduced in Section \ref{sec:per}
and neglect all graphs containing tadpoles.
We will show that this recipe gives a solution
of the functional equation in D dimensions at the
one-loop level. In order to show this we use the identities at the
tree level as in Section \ref{sec:tree} on the integrands.
The examples of the present Section and of Section \ref{sec:two}
demonstrate that indeed up to two loops the functional
identities (\ref{del.4}) and (\ref{del.6}) are satisfied
by neglecting the tadpoles. In Section \ref{sec:more} it
is shown that for higher loop amplitudes tha proof of the
validity of the functional equations is more complex
and necessitate the use of the eq. (\ref{del.4})
at the tree level.
The second important item is how we extract a finite 
part from a divergent amplitude when the limit
$D\to 4$ is taken. We use dimensional subtraction. This procedure
is rather subtle since it might violate the
functional eqs. (\ref{del.4}) and (\ref{del.6}).
The validity of functional equations is
guaranteed if the local counterterms
satisfy eq. (\ref{ren.6}). In particular one has to
fix first the counterterms for the amplitudes involving
only the flat connection and the constrained component $\phi_0$. 
The pole subtraction has
to be performed on the amplitudes normalized according
to eq. (\ref{del.6.1}) as discussed in Section \ref{sec:renor}.
In Section \ref{sec:D=4} we illustrate this strategy.
\par
Let us start from the one-loop
correction of the two-point function of the flat connection.
\subsection{Two-point amplitudes} 
Let us first
consider the one-loop correction to the eq. (\ref{hier.4}). The most
straightforward way to perform this is to {\sl `` close''} two $\phi-$lines
in eq. (\ref{hier.9}). 
From eq. (\ref{tree.6}) we get
\begin{eqnarray}&&
\Gamma^{(1)}_{J_{a_1}^{\mu_1}J_{a'_1}^{\mu'_1}}(p_1) = -i\delta_{a_1 a'_1}
\frac{(2g)^2}{m_D^4}\int \frac{d^D p_2}{(2\pi)^D} 
\frac{(2p_2+p_1)_{\mu_1}(2p_2+p_1)_{\mu'_1}}{p_2^2(p_1+p_2)^2}
\nonumber\\&& 
=
i \frac{(2g)^2}{m_D^4}
\frac{\delta_{a_1 a'_1}}{D-1}
\frac{\Gamma(2-\frac{D}{2})[\Gamma(\frac{D}{2}-1)]^2}
{(4\pi)^\frac{D}{2}\Gamma(D-2)}
\nonumber\\&& 
\frac{i}{(-p_1^2)^{2-\frac{D}{2}}}
(p_1^2g_{\mu_1\mu'_1} - p_{1\mu_1}p_{1\mu'_1})
\nonumber\\&& 
=i\frac{(2g)^2}{m_D^4}
\frac{\delta_{a_1 a'_1}}{D-1}
(p_1^2g_{\mu_1\mu'_1} - p_{1\mu_1}p_{1\mu'_1})I_2(p_1), 
\label{one.1}
\end{eqnarray}
where $I_2$ is defined in eq. (\ref{integral.28}).
With the same procedure one obtains
\begin{eqnarray}&&
\Gamma^{(1)}_{J_{a}^{\mu}\phi_b} =0
\nonumber\\&&
\Gamma^{(1)}_{\phi_a\phi_b} =0.
\label{one.6}
\end{eqnarray}
The vertex functional for the $K_0$ two-point function
is given by 
\begin{eqnarray}&&
i\frac{g^4}{8 m_D^2} \int d^D xd^D yK_0(x)
\Big(\phi^2(x) ~~\phi^2(y)
\Big)_+K_0(y)
\nonumber\\&&
=-i\frac{3g^4}{4 m_D^2} \int  d^D xd^D yK_0(x)
 I_2(x-y)
K_0(y).
\label{one.6.1}
\end{eqnarray}
\subsection{Three-point amplitudes}

\begin{figure}
\epsfxsize=100mm
\centerline{\epsffile{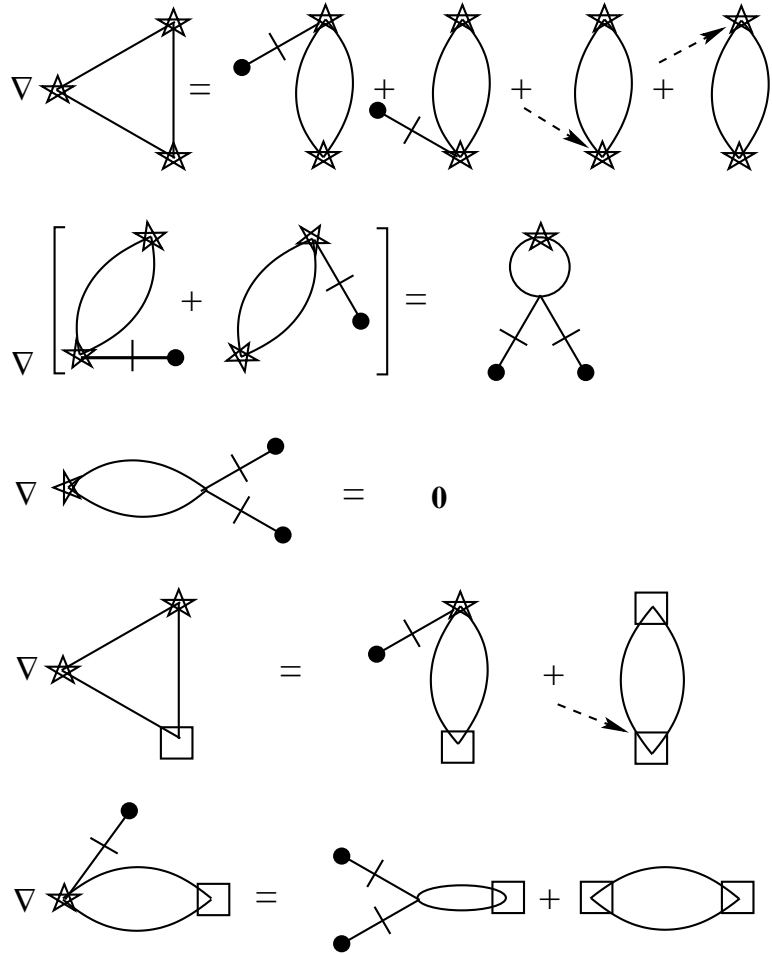}}
\caption{Schematic representation of eq. (\ref{del.6}) 
for the three-point functions in the one-loop approximation.
The meaning of the symbols is: 
$\nabla=i p_1^{\mu_1}$ ,
star = flat-connection, circle = $\phi_a$, $\Box$ = $\phi_0$, dashed arrow =
flow of external momentum, cut line = amputated $\phi_a$-propagator.}
\label{one-loop-3}
\end{figure}

The set of graphs involved at one loop level
for the three-point amplitudes are given in Fig. \ref{one-loop-3}.
Again we follow the hierarchy
of the set of eqs. (\ref{hier.6}-\ref{hier.8}).
The four-divergence on the vertex functions generates some
tadpole graphs. By putting them equal zero the equation
(\ref{del.6}) is satisfied. We shall illustrate this
at the level of integrand since we work in $D$ dimensions.
The starting point is then the three-point function
of the flat-connection. 
\begin{eqnarray}&&
\Gamma^{(1)}_{J_{a_1}^{\mu_1} J_{a_2}^{\mu_2}J_{a_3}^{\mu_3}}(p_1,p_2,p_3)=  
i
\int \frac{d^D k}{(2\pi)^D}
\Gamma^{(0)}_{J_{a_1}^{\mu_1} \phi_{b}\phi_{c}}(p_1,k,-p_1-k)
\nonumber\\&&
\Gamma^{(0)}_{J_{a_2}^{\mu_2} \phi_{c}\phi_{c'}}(p_2,p_1+k,p_3-k)
\Gamma^{(0)}_{J_{a_3}^{\mu_3} \phi_{c'}\phi_{b}}(p_3,-p_3+k,-k)
\nonumber\\&&
\frac{1}{k^2 (p_1+k)^2 (p_3-k)^2}
\label{one.8}
\end{eqnarray}
\par
Now we check the eq. (\ref{tree.3.1})
\begin{eqnarray}&&  
-i
\frac{m_D^2}{2} p^{\mu_1}
\Gamma^{(1)}_{J_{a_1}^{\mu_1} J_{a_2}^{\mu_2}J_{a_3}^{\mu_3}}(p_1,p_2,p_3)
\nonumber\\&&
=
-\frac{i}{2}
\frac{g}{(2\pi)^D}\int d^D k 
\epsilon_{a_1b c} 
\left [
\frac{1}{(p_1+k)^2  (p_3-k)^2}
-
\frac{1}{k^2 (p_3-k)^2 }
\right]
\nonumber\\&&
\frac{(2 ig)^2}{m_D^4}(\delta_{a_3 c}\delta_{a_2b}-
\delta_{a_2 a_3}\delta_{bc})
(p_1-p_3+2k)_{\mu_2}(-p_3+2k)_{\mu_3} 
\nonumber\\&&
= \frac{i}{2}
\frac{(2g)^3}{m_D^4}
\int \frac{d^D k}{(2\pi)^D} \,
\epsilon_{a_1a_2a_3 } 
\left [
\frac{1}{(p_1+k)^2  (p_3-k)^2}
-
\frac{1}{k^2 (p_3-k)^2 }
\right]
\nonumber\\&&
(p_1-p_3+2k)_{\mu_2}(-p_3+2k)_{\mu_3}
\nonumber\\&&
= \frac{i}{2}
\frac{(2 g)^3}{m_D^4}
\int \frac{d^D k}{(2\pi)^D} \,
\epsilon_{a_1a_2a_3 } 
\left [
\frac{(2k-p_2+p_1)_{\mu_2}(2k-p_2)_{\mu_3}}{k^2 (p_2-k)^2}\right.
\nonumber\\&&\left.
-
\frac{(2k-p_3+p_1)_{\mu_2}(2k-p_3)_{\mu_3}}{k^2 (p_3-k)^2 }
\right].
\label{one.9}
\end{eqnarray}
The equation (\ref{hier.8}) involve also the expression in
eq. (\ref{one.1}) and 
\begin{eqnarray}&& 
\Gamma^{(1)}_{\phi_{a_1} J_{a_2}^{\mu_2}J_{a_3}^{\mu_3} }(p_1,p_2,p_3)
\nonumber\\&&
= 
\frac{i}{2}
\frac{(2 g)^3}{m_D^5}
\int \frac{d^D k }{(2\pi)^D} \,
\epsilon_{a_1a_2a_3 } 
\Big [
\frac{(2k-p_2)_{\mu_2}(2k-p_2)_{\mu_3}}{ k^2 (p_2-k)^2}
\nonumber\\&&
-
\frac{(2k-p_3)_{\mu_2}(2k-p_3)_{\mu_3}}{ k^2 (p_3-k)^2}
\Big].
\label{one.10}
\end{eqnarray}
By using eq. (\ref{one.1}) we get
\begin{eqnarray}&& 
=-\frac{i}{2}
\frac{(2 g)^3}{m^5}
\frac{1 }{D-1} \,
\epsilon_{a_1a_2a_3 } \Big [
(p_2^2 g_{{\mu_2}{\mu_3}}-p_{2\mu_2}p_{2\mu_3})I_2(p_2)
\nonumber\\&&
-
(p_3^2 g_{{\mu_2}{\mu_3}}-p_{3\mu_2}p_{3\mu_3})I_2(p_3)
\Big]
\label{one.10.1}
\end{eqnarray}
Then one can easily verify eq. (\ref{hier.8}) by using
eq. (\ref{one.1})
\begin{eqnarray}&& 
-i\frac{m_D^2}{2}p_1^{\mu_1} \Gamma_
{ J_{a_1}^{\mu_1} J_{a_2}^{\mu_2} J_{a_3}^{\mu_3}}
+ m_D\,\,\Gamma_
{ \phi_{a_1} J_{a_2}^{\mu_2} J_{a_3}^{\mu_3}}
\nonumber\\&&
+2g\epsilon_{{a_1}b{a_3}} 
\Gamma_{ J_{a_2}^{\mu_2} J_{b}^{\mu_3}}(p_2)
+2g\epsilon_{{a_1}b{a_2}} 
\Gamma_{ J_{b}^{\mu_2} J_{a_3}^{\mu_3}}(p_3)=0.
\label{one.11}
\end{eqnarray}
\par
Now we consider eq. (\ref{hier.7}). We need
the one loop correction
$\Gamma^{(1)}_{J_{a_1}\phi_{a_2}\phi_{a_3}}$:  
\begin{eqnarray}&& 
\Gamma^{(1)}_{\phi_{a_1} J_{a_2}^{\mu_2} \phi_{a_3}}(p_1,p_2,p_3)
\nonumber\\&&
= 
\frac{ g^3}{m_D^4}
\int \frac{d^D k}{(2\pi)^D}\,
\epsilon_{a_1a_2a_3 } 
(p_1-p_3)^{\mu_1}
\Big [
\frac{(2k+p_2)_{\mu_1}(2k+p_2)_{\mu_2}}{ k^2 (p_2+k)^2}
\Big]
\label{one.12}
\end{eqnarray}
and again by neglecting tadpole graphs the functional
equation (\ref{hier.7}) is satisfied.
Equation (\ref{hier.6}) is simply
\begin{eqnarray}
p_1^\mu
\Gamma^{(1)}_{J_{a_1}^\mu \phi_{a_2}\phi_{a_3}}= 0.
\label{one.13}
\end{eqnarray}
We evaluate  the amplitude $\Gamma^{(1)}_{J\phi\phi}$
in coordinate space
\begin{eqnarray}
i \frac{g^3}{m_D^4} \int d^Dx \int d^D y J_{a_1}^\mu(x)
\Big (\epsilon_{a_1 ab}\partial_\mu\phi_a\phi_b(x)~~
\phi_c\partial_\nu\phi_c\phi_d\partial^\nu\phi_d(y)
\Big)_+
\label{one.13.0.1}
\end{eqnarray}
The relevant vertex function is
\begin{eqnarray}&&
\Gamma^{(1)}_{J\phi\phi}=
-i\frac{g^3}{m_D^4} \int d^Dx \int d^D y J_{c}^\mu(x)\epsilon_{abc}
\nonumber\\&& 
\frac{1}{(D-1)}(\Box g_{\mu\nu}-\partial_\mu
\partial_\nu)I_2(x-y)~~\phi_a\partial^\nu\phi_b(y).
\label{one.13.0.3}
\end{eqnarray}
\par
Finally we have the last equation (\ref{hier.8.1}) 
involving three-point functions. Thus we need
\begin{eqnarray}&&
\Gamma^{(1)}_{J_{a_1}^{\mu_1}\phi_{a_2}K_0}(p_1,p_2,p_3) 
=
 \frac{g^4}{m_D^4}
\int \frac{d^D k}{(2\pi)^D}
\nonumber\\&& 
(3p_1+4 p_2)_{\mu_1}
\frac{1}{k^2(p_3-k)^2},
\label{one.13.1}
\end{eqnarray}
which we now evaluate in coordinate space:
\begin{eqnarray}&&
i\frac{g^4}{2m_D^4} \int d^Dx \int d^D y 
\Big[\Big(2 J_{a_1}^\mu \phi^2 (x)\partial_\mu\phi_{a_1}(x)
\nonumber\\&&
+
\partial_\mu J_{a_1}^\mu \phi^2 (x)\phi_{a_1}(x)\Big) ~~
\phi^2(y)
\Big]_+K_0(y)
\label{one.13.1.1}
\end{eqnarray}
The final result is
\begin{eqnarray}&&
i\frac{g^4}{2m_D^4} \int d^Dx \int d^D y 
\Big[-12 J_{a_1}^\mu\partial_\mu\phi_{a_1}(x)
+4\partial_\mu(J_{a_1}^\mu\phi_{a_1})
\nonumber\\&&
-10
\partial_\mu J_{a_1}^\mu \phi_{a_1}(x)
\Big]I_2(x-y)K_0(y)
\nonumber\\&&
=
-i\frac{g^4}{m_D^4} \int d^Dx \int d^D y 
\Big[4 J_{a_1}^\mu\partial_\mu\phi_{a_1}
+3
\partial_\mu J_{a_1}^\mu \phi_{a_1}
\Big](x)
\nonumber\\&&
I_2(x-y)K_0(y).
\label{one.13.1.3}
\end{eqnarray}
We need a further amplitude
\begin{eqnarray}&&
\Gamma^{(1)}_{\phi_{a_1}\phi_{a_2}K_0}(p_1,p_2,p_3) 
=
i
\frac{g^4}{4m_D^3}
\int \frac{d^D k}{(2\pi)^D}
\nonumber\\&& 
(4(p_1^2+p_2^2)+6p_1p_2 -2kp_3)
\frac{1}{k^2(p_3-k)^2}
\nonumber\\&& 
=
i
\frac{g^4}{4m_D^3}
\int \frac{d^D k}{(2\pi)^D}
(3(p_1^2+p_2^2)+4p_1p_2)
\frac{1}{k^2(p_3-k)^2}.
\label{one.13.3}
\end{eqnarray}
We explicitly evaluate the vertex functional relevant
for $\Gamma^{(1)}_{K_0\phi\phi}$. We need the contractions
on 
\begin{eqnarray}
-i\frac{g^4}{4 m_D^3} \int d^D x d^D y K_0(x)
\Big(\phi(x)^2 ~~\phi_a\partial^\mu\phi_a
\phi_b\partial_\mu\phi_b(y)
\Big)_+.
\label{one.13.3.1}
\end{eqnarray}
The relevant functional is 
\begin{eqnarray}&&
-i\frac{g^4}{4 m_D^3} \int d^D x d^D y K_0(x)
I_2(x-y)\Big(3\Box(\phi^2)(y)
-2\partial^\mu\phi_b\partial_\mu\phi_b(y)
\Big)
\nonumber\\&&
\label{one.13.3.4}
\end{eqnarray}
which together with the amplitudes in eqs. (\ref{one.13.1.3})
and (\ref{one.6.1}) satisfy eq. (\ref{del.6}).
%
\subsection{Four-point amplitudes}

It is of some interest to evaluate some four-point amplitudes,
since on this point the first serious difficulties of the
nonlinear sigma model have been shown \cite{Tataru:1975ys},
\cite{Appelquist:1980ae}. 
We perform the complete evaluation of the amplitudes involving
up to one flat connection insertion. 
We use the same
recipe as before(consisting in the use of tree level Feynman rules
and in neglecting tadpole graphs) and  we evaluate the amplitudes 
$\Gamma_{J\phi\phi\phi}$ and $\Gamma_{\phi\phi\phi\phi}$.
We find convenient to work in coordinate space. From this results
can get also the counterterms and thus check eq. (\ref{ren.4}). 
\subsection{$\Gamma_{J\phi\phi\phi}$}

We perform the necessary contractions on the second
order perturbative expansion of
\begin{eqnarray}
-i\frac{g^4}{2 m_D^5}\int d^Dx \int d^D y \Big[(2
J_{a}^\mu
\phi^2  \partial_\mu\phi_a+\partial_\mu J_{a}^\mu\phi^2\phi_a)(x)
(\phi_c\partial^\nu \phi_c\phi_d\partial_\nu\phi_d)(y)
\Big]_+.
\nonumber\\&& 
\label{one.14.1}
\end{eqnarray}
After some lengthy and straightforward algebra we get
\begin{eqnarray}
\Gamma^{(1)}[J\phi\phi\phi]
&&
-i\frac{g^4}{ m_D^5}\int d^Dx \int d^D y 
\Big\{
\frac{1}{2}
J_a^\mu\partial_\mu\phi_a (x)I_2(x-y)~~ 
\Box\phi^2(y)
\nonumber\\&&
-
J_a^\mu\partial_\mu\phi_a(x) I_2(x-y)~~ \partial_\nu\phi_d\partial^\nu\phi_d(y)
\nonumber\\&&
+ J_a^\mu  \phi_a(x) I_2(x-y)
 ~~\partial_\mu(\partial_\nu \phi_d \partial^\nu\phi_d)(y)
\nonumber\\&& 
-\frac{3}{2} J_a^\mu \phi_a (x)I_2(x-y)
  \partial_\mu\Box\phi^2(y)
\nonumber\\&&
-\frac{2}{(D-1)}J_a^\mu \phi_c(x)I_2(x-y)
 ~~\left(\Box g_{\mu\nu}
-
\partial_\mu\partial_\nu\right)\left(\partial^\nu \phi_c \phi_a\right)(y)
\nonumber\\&&
+2J_a^\mu \partial_\mu\phi_c (x)I_2(x-y)
 ~~\partial^\nu\phi_a\partial_\nu \phi_c (y)
\Big\}
\nonumber\\&&
\label{one.14.4}
\end{eqnarray}
Now we take the derivative in order to check eq. (\ref{del.6}).
\begin{eqnarray}&&
\frac{\delta}{\delta J_a^\mu}
\Gamma^{(1)}[J\phi\phi\phi] =
-i\frac{g^4}{ m_D^5}\int d^D y 
~~\Big\{
\frac{1}{2}\Box\phi_a I_2(x-y)\Box\phi^2
+\frac{1}{2}\partial^\mu\phi_a I_2(x-y)
  \partial_\mu\Box\phi^2
\nonumber\\&& 
- \Box\phi_a I_2(x-y)
 ~~\partial_\nu \phi_d \partial^\nu\phi_d
+  \phi_a I_2(x-y)
 ~~\Box(\partial_\nu \phi_d \partial^\nu\phi_d)
\nonumber\\&&
 -\frac{3}{2}\partial^\mu\phi_a I_2(x-y)
  \partial_\mu\Box\phi^2
 -\frac{3}{2}\phi_a I_2(x-y)
 \Box^2\phi^2
\nonumber\\&& 
-\frac{2}{(D-1)}\partial^\mu\phi_cI_2(x-y)
 ~~\left(\Box g_{\mu\nu}
-\partial_\mu\partial_\nu\right)\left(\partial^\nu \phi_c \phi_a\right)
\nonumber\\&& 
+2 \Box\phi_c I_2(x-y)
 ~~\partial^\nu\phi_a\partial_\nu \phi_c 
+2 \partial^\mu\phi_c I_2(x-y)
 ~~\partial_\mu(\partial^\nu\phi_a\partial_\nu \phi_c )
\Big\}.
\nonumber\\&& 
\label{one.14.5}
\end{eqnarray}
i.e. we can obtain the first term in eq. (\ref{del.6})
\begin{eqnarray}&& 
\frac{m_D^2}{2}
\partial^\mu\frac{\delta}{\delta J_a^\mu }\Gamma_{J\phi\phi\phi}=
-i
\frac{g^4}{2 m_D^3}~
\int d^D y 
\Big\{
\frac{1}{2}\Box\phi_a I_2(x-y)
\Box\phi^2
\nonumber\\&&
- \Box\phi_a I_2(x-y)
 ~~\partial_\nu \phi_d \partial^\nu\phi_d
+  \phi_a I_2(x-y)
 ~~\Box(\partial_\nu \phi_d \partial^\nu\phi_d)
\nonumber\\&&
 -\partial^\mu\phi_a I_2(x-y)
  \partial_\mu\Box\phi^2
 -\frac{3}{2}\phi_a I_2(x-y)
 \Box^2\phi^2
\nonumber\\&& 
-\frac{2}{(D-1)}\partial^\mu\phi_cI_2(x-y)
 ~~\left(\Box g_{\mu\nu}
-\partial_\mu\partial_\nu\right)\left(\partial^\nu \phi_c \phi_a\right)
\nonumber\\&& 
+2 \Box\phi_c I_2(x-y)
 ~~\partial^\nu\phi_a\partial_\nu \phi_c 
+2 \partial^\mu\phi_c I_2(x-y)
 ~~\partial_\mu(\partial^\nu\phi_a\partial_\nu \phi_c )
\Big\}.
\nonumber\\&& 
\label{one.14.6}
\end{eqnarray}
\subsection{$\Gamma_{\phi\phi\phi\phi}$}

We perform the necessary contractions on the second
order perturbative expansion of
\begin{eqnarray}
i\frac{1}{2}\left(\frac{g^2}{2m_D^2}\right)^2\int d^Dx~d^D y 
\left(
\phi_a\partial^\mu\phi_a\phi_b\partial_\mu\phi_b(x)~
\phi_c\partial^\nu\phi_c\phi_d\partial_\nu\phi_d(y)
\right).
\label{one.45}
\end{eqnarray}
After some algebra one gets
\begin{eqnarray}&&
i\left(\frac{g^2}{2m_D^2}\right)^2\int d^Dx~d^D y 
\nonumber\\&&
\Big\{-\frac{3}{4}\Box\phi^2(x) ~I_2(x-y)~\Box\phi^2 (y)
+\Box\phi^2(x)
~I_2(x-y)~\partial^\nu\phi_c\partial_\nu\phi_c(y)
\nonumber\\&&
-\partial^\mu\phi_c\partial_\mu\phi_b (x)
~I_2(x-y)~\partial^\nu\phi_c\partial_\nu\phi_b(y)
\nonumber\\&&
-\frac{1}{(D-1)} \partial^\mu\phi_a\phi_b(x)~~
(\Box g_{\mu\nu}-\partial_\mu\partial_\nu)~I_2(x-y)~ 
\partial_\nu\phi_a\phi_b (y)
\Big\}.
\nonumber\\&&
\label{one.46}
\end{eqnarray}
The singular part at $D=4$ of the expression in eq. (\ref{one.46})
agrees with the results of Reference \cite{Appelquist:1980ae}.
One simply uses the identities%
\begin{eqnarray}&&
\Box\phi^2 ~\Box\phi^2=4[(\phi_a\Box\phi_a)^2
+2\phi_a\Box\phi_a\partial_\mu\phi_b\partial^\mu\phi_b 
+\partial_\nu\phi_a\partial^\nu\phi_a
\partial_\mu\phi_b\partial^\mu\phi_b]
\nonumber\\&&
\Box\phi^2\partial_\nu\phi_a\partial^\nu\phi_a
=2[\partial_\nu\phi_a\partial^\nu\phi_a
\partial_\mu\phi_b\partial^\mu\phi_b 
+\phi_a\Box\phi_a\partial_\mu\phi_b\partial^\mu\phi_b  ]
\nonumber\\&&
\int d^Dx
\partial^\mu\phi_a\phi_b~~
(\Box g_{\mu\nu}-\partial_\mu\partial_\nu)~(\partial^\nu\phi_a\phi_b )
\nonumber\\&&
= 
-
\int d^Dx[\partial_\nu(\partial^\mu\phi_a\phi_b)
\partial^\nu(\partial_\mu\phi_a\phi_b)
-\partial_\nu(\partial^\mu\phi_a\phi_b)
\partial_\mu(\partial^\nu\phi_a\phi_b)]
\nonumber\\&&
= 
\int d^Dx[\partial_\nu\phi_a\partial^\nu\phi_b
\partial_\mu\phi_a\partial^\mu\phi_b
-\partial_\nu\phi_a\partial^\nu\phi_a
\partial_\mu\phi_b\partial^\mu\phi_b].
\label{one.46app}
\end{eqnarray}
\par\noindent
In order to check eq. (\ref{del.6}) we take the derivative 
with respect to $\phi_a$
\begin{eqnarray}&&
m_D\Gamma_{\phi_a}=
i\frac{g^4}{4m_D^3}\int~d^D y 
\Big\{
-3\phi_a ~I_2(x-y)~\Box^2\phi^2
\nonumber\\&&
+2\phi_a
~I_2(x-y)~\Box(\partial^\nu\phi_c\partial_\nu\phi_c)
-2\Box\phi_a I_2(x-y)\Box\phi^2
-2\partial_\nu\phi_a I_2(x-y)\partial^\nu\Box\phi^2
\nonumber\\&&
+4\Box\phi_b 
~I_2(x-y)~\partial^\nu\phi_a\partial_\nu\phi_b
+4\partial_\mu\phi_b 
~I_2(x-y)~\partial^\mu(\partial^\nu\phi_a\partial_\nu\phi_b)
\nonumber\\&&
-\frac{2}{(D-1)} \partial^\mu\phi_c~~
(\Box g_{\mu\nu}-\partial_\mu\partial_\nu)~I_2(x-y)~ 
(\partial_\nu\phi_c\phi_a )
\nonumber\\&&
+\frac{2}{(D-1)} \partial^\mu\phi_c~~
(\Box g_{\mu\nu}-\partial_\mu\partial_\nu)~I_2(x-y)~ 
[\partial_\nu(\phi_a\phi_c )- \phi_a\partial_\nu\phi_c ]
\Big\}
\nonumber\\&&
=
i\frac{g^4}{4m_D^3}\int~d^D y 
\Big\{
-3\phi_a ~I_2(x-y)~\Box^2\phi^2
\nonumber\\&&
+2\phi_a
~I_2(x-y)~\Box(\partial^\nu\phi_c\partial_\nu\phi_c)
-2\Box\phi_a I_2(x-y)\Box\phi^2
-2\partial_\nu\phi_a I_2(x-y)\partial^\nu\Box\phi^2
\nonumber\\&&
+4\Box\phi_b 
~I_2(x-y)~\partial^\nu\phi_a\partial_\nu\phi_b
+4\partial_\mu\phi_b 
~I_2(x-y)~\partial^\mu(\partial^\nu\phi_a\partial_\nu\phi_b)
\nonumber\\&&
-\frac{4}{(D-1)} \partial^\mu\phi_c~~
(\Box g_{\mu\nu}-\partial_\mu\partial_\nu)~I_2(x-y)~ 
(\partial_\nu\phi_c\phi_a )
\Big\}.
\nonumber\\&&
\label{one.46.1.0}
\end{eqnarray}
The quantities in eqs. (\ref{one.14.6}), (\ref{one.46.1.0}) and 
(\ref{one.13.3.4}) do satisfy eq. (\ref{del.6}).
%
%

\section{Two-loops}
\label{sec:two}
Now we consider some cases at the two-loop level.
We use the Feynman rules derived  in Sec. \ref{sec:tree}.

\subsection{Two-point functions}

\begin{figure}
\epsfxsize=100mm
\centerline{\epsffile{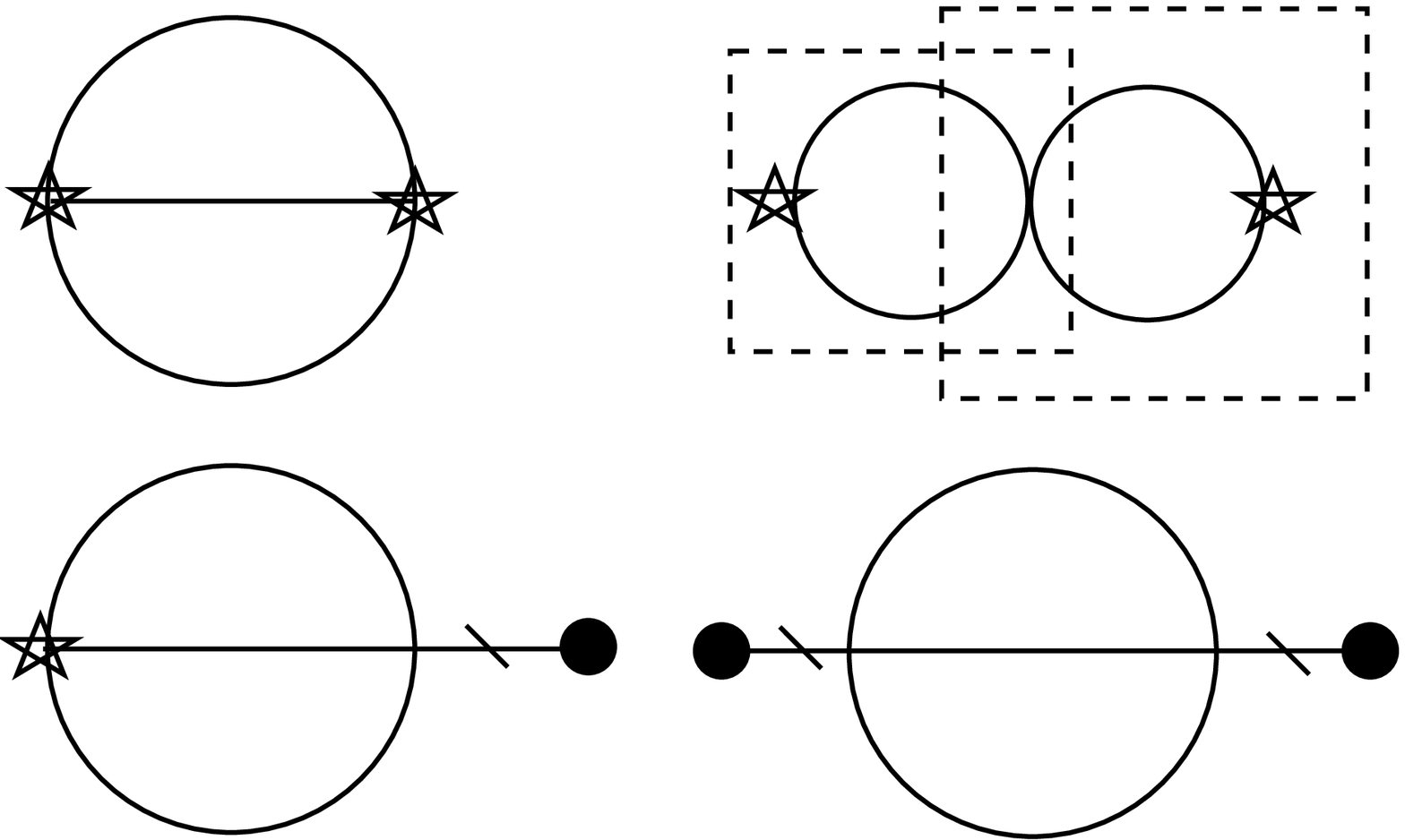}}
\caption{Two-loop graphs for the two-point functions.}
\label{two-point-2}
\end{figure}

We consider the contribution of the graphs depicted in
Fig. \ref{two-point-2}
For the evaluation of the $\phi-\phi$ two-point function
we use the  Feynman rule in eq. (\ref{tree.14}).
\begin{eqnarray}&&
\Gamma^{(2)}_{\phi_{a_1} \phi_{a'_1} }(p_1)
=
i\frac{g^4}{6m_D^4} \int
\frac{ d^D p_2}{(2\pi)^{D}}\frac{ d^D p_3}{(2\pi)^{D}}
\frac{i^3}{p_2^2p_3^2p_4^2} 
\nonumber\\&& 
\Big( \delta_{a_3a_4}\delta_{a_1a_2}( p_2+p_1)^2
\nonumber\\&& +
\delta_{a_2a_4}\delta_{a_1a_3}( p_3+p_1)^2 +
\delta_{a_2a_3}\delta_{a_1a_4}( p_4+p_1)^2
\Big) 
\nonumber\\&& 
\Big( \delta_{a_3a_4}\delta_{a'_1a_2}( p_2+p_1)^2
\nonumber\\&& +
\delta_{a_2a_4}\delta_{a'_1a_3}( p_3+p_1)^2 +
\delta_{a_2a_3}\delta_{a'_1a_4}( p_4+p_1)^2
\Big)
\nonumber\\&& 
=\frac{g^4}{6m_D^4} \delta_{a_1a'_1}\int
\frac{ d^D p_2}{(2\pi)^{D}} \frac{ d^D p_3}{(2\pi)^{D}}
\frac{1}{p_2^2p_3^2p_4^2} 
\nonumber\\&& 
\left[
9[(p_1+p_2)^2]^2 + 6 (p_1+p_2)^2(p_1+p_3)^2
\right]
\nonumber\\&& 
=\frac{g^4}{6m_D^4}  \delta_{a_1a'_1}\int
\frac{ d^D p_2}{(2\pi)^{D}} \frac{ d^D p_3}{(2\pi)^{D}}
\frac{1}{p_2^2p_3^2p_4^2}  
\left\{
9[p_1^4+4(p_1p_2)^2 \right.
\nonumber\\&&  
\left.
+ 4p_1^2(p_1p_2)]
+ 6 [p_1^4+ 2p_1^2(p_1p_3)
+2p_1^2(p_1p_2) + 4 (p_1p_2)(p_1p_3)]
\right\}
\nonumber\\&& 
\label{two-loop.1}
\end{eqnarray}
The integration over $p_3$ allows to use eq. (\ref{integral.32})
\begin{eqnarray}&&
\Gamma^{(2)}_{\phi_{a_1} \phi_{a'_1} }(p_1)
= \frac{g^4}{6m_D^4}\delta_{a_1a'_1} \langle\left[
15p_1^4 + 48 p_1^2(p_1p_2)+24(p_1p_2)^2
\right]\rangle_3
\nonumber\\&& 
= \frac{g^4}{6m_D^4}\delta_{a_1a'_1}
\frac{5D-4}{3D-4}I_3(p_1)p_1^4.
\label{two-loop.2}
\end{eqnarray}
$I_3$ is given in eq. (\ref{integral.38}).
Now we evaluate the next relevant two-point amplitude
(the second graph in Figure \ref{two-point-2} is not relevant
for the check of eq. (\ref{hier.5}) since it is proportional
to a pure transverse tensor)
\begin{eqnarray}&&
\Gamma^{(2)}_{J_{a_1}^{\mu_1} J_{a'_1}^{\mu'_1} }(p_1)= 
\frac{i}{6}
\int  \frac{d^D p_2}{(2\pi)^{2D}}\int \frac{d^D p_3}{(2\pi)^{2D}}
\nonumber\\&&
\Gamma^{(0)}_
{ J_{a_1}^{\mu_1} \phi_{a_2} \phi_{a_3} \phi_{a_4}}
(p_1,p_2,p_3,p_4)
\frac{i^3}{p_2^2p_3^2p_4^2}\left .
\Gamma^{(0)}_
{ J_{a'_1}^{\mu'_1} \phi_{a_2} \phi_{a_3} \phi_{a_4}}
(p'_1,-p_2,-p_3,-p_4)
\right|_{p_4=-p_1-p_2-p_3}
\nonumber\\&&
=
\frac{1}{6}\left(
\frac{-2ig^2}{m_D^3}
\right)^2
\int  \frac{d^D p_2}{(2\pi)^{2D}}\int \frac{d^D p_3}{(2\pi)^{2D}}
\frac{1}{p_2^2p_3^2p_4^2}
\Big( \delta_{a_3a_4}\delta_{a_1a_2}( 2p_2+p_1)
\nonumber\\&& +
\delta_{a_2a_4}\delta_{a_1a_3}(2 p_3+p_1) +
\delta_{a_2a_3}\delta_{a_1a_4}(2 p_4+p_1)
\Big)_\mu
\nonumber\\&& 
\Big( \delta_{a_3a_4}\delta_{a'_1a_2}(-2 p_2+p'_1)
\nonumber\\&& +
\delta_{a_2a_4}\delta_{a'_1a_3}(-2 p_3+p'_1) +
\delta_{a_2a_3}\delta_{a'_1a_4}(-2 p_4+p'_1)
\Big)_{\mu'}
\nonumber\\&&
=\frac{1}{6} \left(
\frac{2g^2}{m_D^3}
\right)^2
\int  \frac{d^D p_2}{(2\pi)^{2D}}\int \frac{d^D p_3}{(2\pi)^{2D}}
\frac{1}{p_2^2p_3^2p_4^2}
\Big(9( 2p_2+p_1)_\mu( 2p_2+p_1)_{\mu'}
\nonumber\\&&
+6( 2p_2+p_1)_\mu( 2p_3+p_1)_{\mu'}
\Big)\delta_{a_1a'_1}
.
\label{two-loop.3}
\end{eqnarray}
We use the equation (\ref{integral.32}) for the integration over $p_3$
\begin{eqnarray}&&
\Gamma^{(2)}_{J_{a_1}^{\mu_1} J_{a'_1}^{\mu'_1} }(p_1)
\nonumber\\&&
=
\frac{1}{6} \left(
\frac{2g^2}{m_D^3}
\right)^2
\int  \frac{d^D p_2}{(2\pi)^{2D}}\int \frac{d^D p_3}{(2\pi)^{2D}}
\frac{1}{p_2^2p_3^2p_4^2}
\Big(
9( 2p_2+p_1)_\mu( 2p_2+p_1)_{\mu'}
\nonumber\\&&
+6( 2p_2+p_1)_\mu( 2[-\frac{1}{2}(p_1+p_2)]+p_1)_{\mu'}
\Big)
\delta_{a_1a'_1}
\nonumber\\&&
=
\frac{1}{6} \left(
\frac{2g^2}{m_D^3}
\right)^2
\int  \frac{d^D p_2}{(2\pi)^{2D}}\int \frac{d^D p_3}{(2\pi)^{2D}}
\frac{1}{p_2^2p_3^2p_4^2}
\Big(
9p_{1\mu}p_{1\mu'}
\nonumber\\&&
+18p_{2\mu}p_{1\mu'}
+12p_{1\mu}p_{2\mu'}+p_{2\mu}p_{2\mu'}
\Big)
\delta_{a_1a'_1}
\nonumber\\&&
=
\frac{4}{6}\delta_{a_1a'_1} \left(
\frac{g^2}{m_D^3}
\right)^2 I_3(p_1)
\Big(
-p_{1\mu}p_{1\mu'}+\frac{8}{3D-4}(-p_1^2g_{\mu\mu'}+Dp_{1\mu}p_{1\mu'})
\Big)
\nonumber\\&&
=
\frac{2}{3}\delta_{a_1a'_1} \left(
\frac{g^2}{m_D^3}
\right)^2 \frac{I_3(p_1)}{3D-4}
\Big(
-8p_1^2g_{\mu\mu'}+(5D+4)p_{1\mu}p_{1\mu'}
\Big).
\label{two-loop.4}
\end{eqnarray}
Now we can check the validity of eqs. (\ref{hier.4}) and (\ref{hier.5})
\begin{eqnarray}
m_D^2
\partial^\mu \partial^{\mu'}\Gamma_{ J_{a_1}^\mu J_{a'_1}^{\mu'}}
-4\,\Gamma_{ \phi_{a_1}\phi_{a'_1}} =0.
\label{two-loop.5}
\end{eqnarray}
\subsection{Three-point functions}

\begin{figure}
\epsfxsize=100mm
\centerline{\epsffile{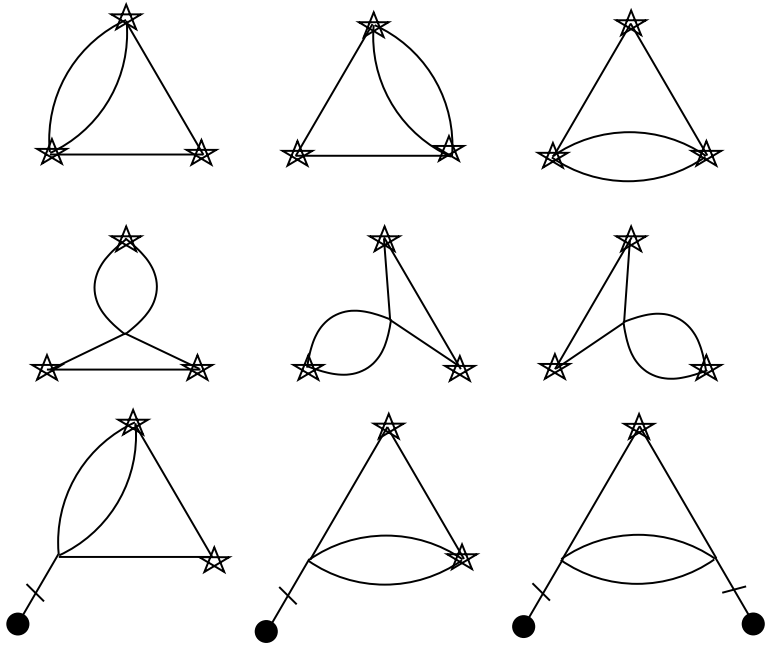}}
\caption{Two-loop graphs for the three-point functions.}
\label{two-point-3}
\end{figure}

\begin{figure}
\epsfxsize=100mm
\centerline{\epsffile{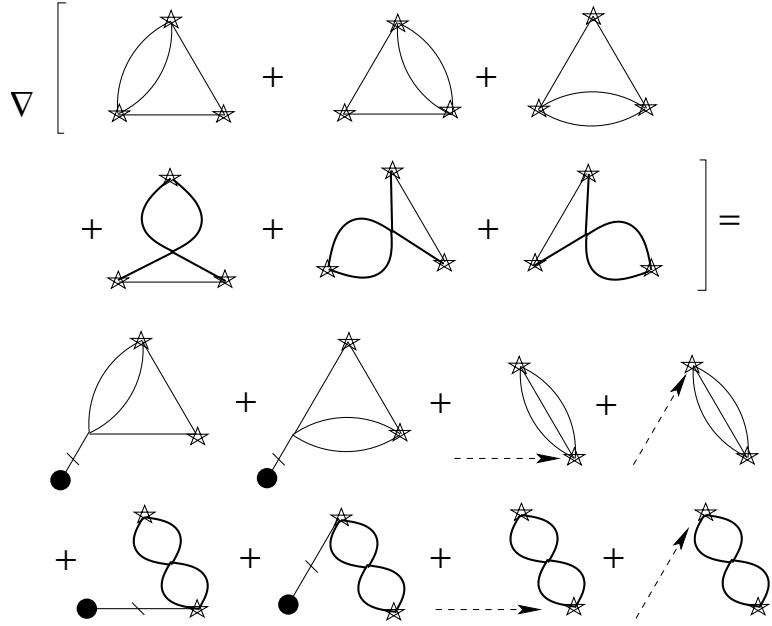}}
\caption{Graphical representation of eq. (\ref{hier.8}) at
two loop level.  }
\label{decoration-3}
\end{figure}

The two-loop three-point functions are drawn in Fig. \ref{two-point-3}.
When we try to check the validity of the relation in eq. (\ref{hier.8})
we get again some tadpole graphs that 
we put equal zero. Thus we are left only with the graphs in 
Figure \ref{decoration-3}. Eqs. (\ref{tree.3.1}) and (\ref{tree.9})
allow to prove the validity of the functional equation (\ref{hier.8}).
\par
Similar arguments can be used in order to prove eqs. (\ref{hier.6})
and (\ref{hier.7}) for the rest of the amplitudes in Figure \ref{two-point-3}.
\section{More loops}
\label{sec:more}
Here we discuss the r\"ole of the equations of motion
eq. (\ref{del.4p}) in the proof of the validity of
eq. (\ref{del.6}). The tree level solutions of eq. (\ref{del.4})
can be readily constructed from the Feynman vertices given by
$\Gamma^{(0)}$ in eq. (\ref{ren.2}) by the standard procedure
recalled in Section \ref{sec:rec}.
\par
The crucial point of our approach is the statement that na{\"\i}ve
Feynman rules given by $\Gamma^{(0)}$ in eq. (\ref{ren.2})
yield amplitudes that are solutions of the functional equations
(\ref{del.6}) for generic D dimensions. Up to two loops
the property of dimensional regularization (\ref{ren.3})
\begin{eqnarray}
\int d^Dk \frac{1}{(k^2)^\alpha}=0,\qquad \alpha \in {\cal C}
\label{ren.3p}
\end{eqnarray}
implies 
the validity of the functional equations (\ref{del.6})
at the perturbative level by imposing that all tadpole graphs are 
zero.
\par
\begin{figure}
\epsfxsize=100mm
\centerline{\epsffile{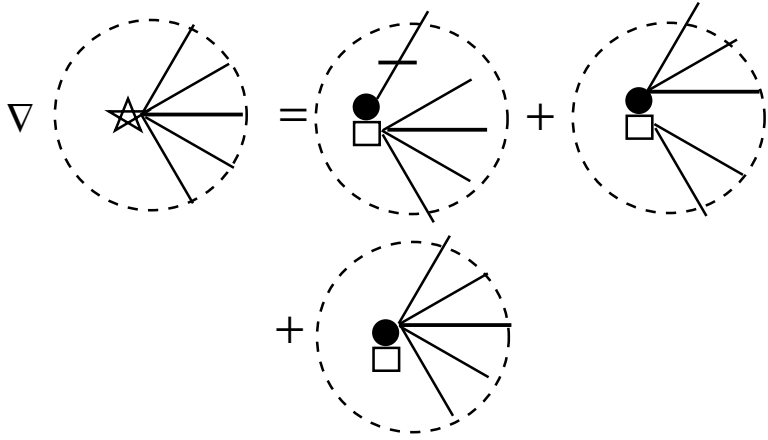}}
\caption{Perturbative representation of the functional identity}
\label{ward}
\end{figure}
\begin{figure}
\epsfxsize=100mm
\centerline{\epsffile{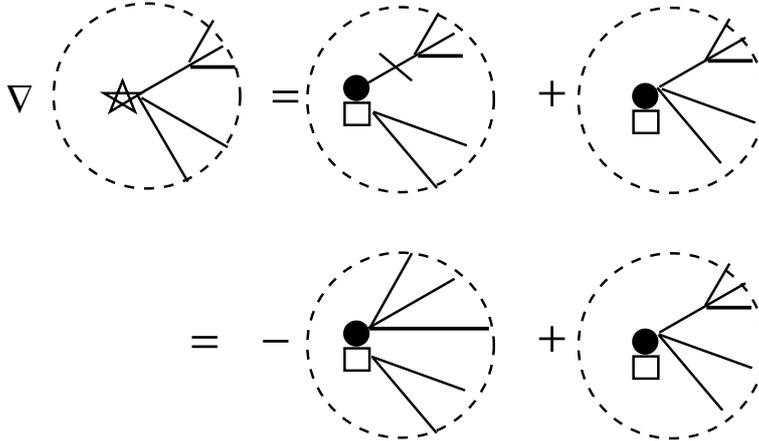}}
\caption{Perturbative check of functional identity}
\label{ward-1}
\end{figure}

For higher number of loops one needs the equations of motion
in (\ref{del.4p}). We illustrate this fact for the four-loop
corrections of the amplitude $\Gamma_{J\phi}$. The equation
to be proved is (\ref{hier.4}). From the symmetry $\phi_a\to -\phi_a$
we see that starting from $F_a^\mu$ we have three or five $\phi$
lines as depicted in Figures \ref{ward} and \ref{ward-1}.
When one uses the functional identities for these vertices, new terms
appear in comparison with the single one present in eq. (\ref{hier.4}).
In particular (\ref{tree.9})
\begin{eqnarray}
\frac{m_D^2}{2}\partial^{\mu_1}\Gamma^{(0)}_
{ J_{a_1}^{\mu_1} \phi_{a_2} \phi_{a_3}\phi_{a_4}}
+  m_D\,\,\Gamma^{(0)}_
{ \phi_{a_1} \phi_{a_2}\phi_{a_3}\phi_{a_4}}
+\sum_{j=2}^4 
\Gamma^{(0)}_{K_0 \phi_{a_{j+1}}
 \phi_{a_{j+2}}}
\Gamma^{(0)}_
{ \phi_{a_1} \phi_{a_j}}
=0
\nonumber\\&&
\label{tree.9p}
\end{eqnarray}
and
\begin{eqnarray}&&
\frac{m_D^2}{2}\partial^{\mu_1}\Gamma^{(0)}_
{ J_{a_1}^{\mu_1} \phi_{a_2} \phi_{a_3}\phi_{a_4}\phi_{a_5}\phi_{a_6}}
+  m_D\,\,\Gamma^{(0)}_
{ \phi_{a_1} \phi_{a_2}\phi_{a_3}\phi_{a_4}\phi_{a_5}\phi_{a_6}}
\nonumber\\&&
+\sum_{\cal P} 
\Gamma^{(0)}_{K_0  \phi_{{ a'_2}} \phi_{ a'_3}}
\Gamma^{(0)}_
{ \phi_{a_1} \phi_{a'_4}\phi_{a'_5}\phi_{a'_6}}
+\sum_{j=2}^6 
\Gamma^{(0)}_{K_0 \phi_{a_{j+1}}
 \phi_{a_{j+2}}\phi_{a_{j+3}}\phi_{a_{j+4}}}
\Gamma^{(0)}_
{ \phi_{a_1} \phi_{a_j}}
=0
\nonumber\\&&
\label{tree.9pp}
\end{eqnarray}
where $\{a'_2, a'_3,a'_4,a'_5,a'_6\}\in{\cal P}$ is any partition 
into two sets of 2 and 3 elements.
\par
The last element in eq. (\ref{tree.9pp}) gives zero by using eq. 
(\ref{ren.3p}),
while the rest of the unwanted terms cancel by using the equations
of motion at the tree level, as depicted in Figures \ref{ward} 
and \ref{ward-1}. Our conjecture is that this cancellation is present
at any order in the loop expansion and for any amplitude involving
$F_a^\mu$, $\phi_a$ and $\phi_0$.
%
\section{Subtraction procedure at $D=4$}
\label{sec:D=4}
In this Section we provide some examples of the subtraction
procedure in order to define the amplitudes at $D=4$.
For the one loop corrections this amounts to select
the pole part, to construct the counterterm by using
the normalized amplitudes as in eq. (\ref{del.6.1})
and to get the finite part. In the example involving
two loop corrections it will be shown that the recursive
subtraction works correctly since the counterterms
involve only local terms in the vertex functional.
\subsection{One loop counterterms for $\Gamma_{JJ},  ~\Gamma_{K_0K_0}$}
From eq. (\ref{one.1}) we see that the pole part of
the two-point function involving only the flat connection
is
\begin{eqnarray}\left .
\Gamma^{(1)}_{J_{a_1}^{\mu_1}J_{a'_1}^{\mu'_1}}(p_1) \right|_{\rm POLE}
=
\frac{1 }{D-4}
 \frac{g^2}{m^4}
\frac{\delta_{a_1 a'_1}}{6\pi^2}
(p_1^2g_{\mu_1\mu'_1} - p_{1\mu_1}p_{1\mu'_1}).
\label{one.1pole}
\end{eqnarray}
The counterterm has to be evaluated on the
normalized amplitude as in eq. (\ref{del.6.1}):
\begin{eqnarray}
{\widehat\Gamma}^{(1)}[JJ] 
=
\frac{1}{(D-4)}\left(
\frac{m}{m_D}
\right)^{2}
 \frac{g^2}{12\pi^2m^4}
\int d^D x J_{a}^{\mu}
(\Box g_{\mu\nu} - \partial_\mu\partial_\nu) J_{a}^{\nu}.
\label{one.1pole1}
\end{eqnarray}
A similar analysis gives 
the counterterm associated to the amplitude in eq. (\ref{one.6.1})
\begin{eqnarray}
{\widehat\Gamma}^{(1)}[K_0K_0]
=
\frac{1}{(D-4)}\frac{3g^4}{2 m^2 (4\pi)^2} 
\int  d^D x K_0K_0.
\label{one.6.1c}
\end{eqnarray}
%
%
\subsection{One loop counterterms for $\Gamma_{JJJ}, ~\Gamma_{JJ\phi},
~\Gamma_{J\phi\phi}, ~\Gamma_{J\phi K_0}, ~\Gamma_{K_0\phi\phi},~\Gamma_{JJK_0}$}
From eq. (\ref{one.9}) the pole part of the three-point flat
connection $\Gamma_{JJJ}$ is 
\begin{eqnarray}&&
\left .
\Gamma^{(1)}_{J_{a_1}^{\mu_1}
 J_{a_2}^{\mu_2}J_{a_3}^{\mu_3}}(p_1,p_2,p_3)\right|_{\rm POLE}
=
-
\frac{i}{3\pi^2}\left(\frac{g}{m^2}\right)^3
\left(
\frac{1}{D-4}
\right)
\epsilon_{a_1a_2{a_3}}
\nonumber\\&&
\Big[
(p_3-p_2)_{\mu_1}g_{{\mu_2}{\mu_3}}
+
(p_1-p_3)_{\mu_2}g_{{\mu_1}{\mu_3}}
+
(p_2-p_1)_{\mu_3}g_{{\mu_1}{\mu_2}}
\Big]
\label{one.8.6}
\end{eqnarray}
Then the counterterm is
\begin{eqnarray}
{\widehat\Gamma}^{(1)}[JJJ] = \frac{1}{3\pi^2}\left(\frac{g}{m^2}\right)^3
\left(
\frac{1}{D-4}
\right)
\left(
\frac{m}{m_D}
\right)^{4}
\epsilon_{abc}\int d^Dx \partial_\mu J_{a\nu} J_b^\mu J_c^\nu.
\label{one.8.7}
\end{eqnarray}
\par
The pole part of the one loop amplitude $\Gamma_{JJ\phi}$ can be
obtained form eq. (\ref{one.10.1})
\begin{eqnarray}&& \left .
\Gamma^{(1)}_{\phi_{a_1} J_{a_2}^{\mu_2}J_{a_3}^{\mu_3} }(p_1,p_2,p_3)
\right |_{\rm POLE}
= -\frac{1}{6\pi^2}\frac{ g^3}{m^5}\frac{1 }{D-4}
\nonumber\\&&
\epsilon_{a_1a_2a_3 } 
\Big [
(p_2^2 g_{{\mu_2}{\mu_3}}-p_{2\mu_2}p_{2\mu_3})
-
(p_3^2 g_{{\mu_2}{\mu_3}}-p_{3\mu_2}p_{3\mu_3})
\Big],
\label{one.10.2}
\end{eqnarray}
which corresponds to a counterterm (according to eq. (\ref{del.6.1}))
\begin{eqnarray}&&
{\widehat\Gamma}^{(1)}[ JJ\phi]=
\frac{1}{6\pi^2}\frac{1 }{D-4}\frac{ g^3}{m^5}
\left(
\frac{m}{m_D}
\right)^{3}
\epsilon_{a_1a_2a_3 } 
\nonumber\\&&
\int d^Dx \phi_{a_1}J_{a_2}^{\mu_2}(\Box g_{{\mu_2}{\mu_3}}
-\partial_{\mu_2}\partial_{\mu_3})J_{a_3}^{\mu_3}.
\label{one.10.3}
\end{eqnarray}
%
\par
The pole part of the one loop amplitude $\Gamma_{J\phi\phi}$ can be
obtained form eq. (\ref{one.13.0.3}).
The counterterm according to the rule in eq. (\ref{del.6.1})
is
\begin{eqnarray}&&
{\widehat\Gamma}^{(1)}[J\phi\phi]
=
\frac{1}{D-4}\frac{1}{24\pi^2}\frac{g^3}{m^4}
\left(
\frac{m}{m_D}
\right)^{2}
\epsilon_{abc}\int d^Dx (\Box g_{\mu\nu}-\partial_\mu
\partial_\nu)J_{c}^\mu~~\phi_a\partial^\nu\phi_b
.
\nonumber\\
\label{one.13.0.4}
\end{eqnarray}
\par
The pole part of the one loop amplitude $\Gamma_{JK_0\phi}$ can be
obtained form eq. (\ref{one.13.1.3}).
The counterterm according to the rule in eq. (\ref{del.6.1})
is
\begin{eqnarray}&&
{\widehat\Gamma}^{(1)}[JK_0\phi]
=\frac{2}{D-4}\frac{g^4}{m^4(4\pi)^2}\left(
\frac{m}{m_D}
\right)^{2}
\int d^Dx 
\Big[4 J_{a_1}^\mu\partial_\mu\phi_{a_1}
+3
\partial_\mu J_{a_1}^\mu \phi_{a_1}
\Big]K_0.
\nonumber\\
\label{one.13.1.3c}
\end{eqnarray}
\par
The counterterm associated to the amplitude in eq. (\ref{one.13.3.4})
is
\begin{eqnarray}&&
{\widehat\Gamma}^{(1)}[K_0\phi\phi] =
\frac{1}{D-4}
\frac{g^4}{2 m^3(4\pi)^2}\left(
\frac{m}{m_D}
\right) \int d^D x  K_0
\Big(3\Box(\phi^2)
-2\partial^\mu\phi_b\partial_\mu\phi_b
\Big).
\nonumber\\&&
\label{one.13.3.4c}
\end{eqnarray}
\par
By using the equation (\ref{del.6.1}) and the results
already obtained in eqs. (\ref{one.6.1c}), (\ref{one.13.1.3c}) and
(\ref{sigma.13}) we get
\begin{eqnarray}
{\widehat\Gamma}^{(1)}[JJK_0] = \frac{1}{D-4}
\frac{8 g^4}{m^5(4\pi)^2}\left(
\frac{m}{m_D}
\right)^3
\int d^Dx K_0 J^\mu_a J_{a\mu}.
\label{one.6.3p}
\end{eqnarray}

\subsection{A two-loop example: $\Gamma_{JJ}^{(2)}$}
%
The simplest example of two loop amplitude is given
by the graphs in Figure \ref{two-point-2}. The first
graph is somewhat trivial since the subgraph counterterms
are inserted in a tadpole, i.e. they give no contribution.
That is the reason why the amplitudes in eqs. (\ref{two-loop.2})
and (\ref{two-loop.4}) contain no double poles in $D=4$.
In this respect the second graph in Figure \ref{two-point-2}
is very interesting. In the generic dimension $D$
the amplitude is given by
\begin{eqnarray}&&
\Gamma_{J_a^\mu J_b^\nu}^{(2)}
=
\delta_{ab}
\frac{2g^4}{m_D^6} 
\frac{-p^2}{(D-1)^2}(-p^2 g_{\mu\nu}+p_\mu p_\nu)
\nonumber\\&&
\frac{1}{[-p^2]^{3-D}}
\frac{\Gamma\left(2-\frac{D}{2}\right)^2}{(4\pi)^{D}}
\frac{\Gamma\left(\frac{D}{2}-1\right)^4}{\Gamma(D-2)^2}.
\label{two-loop.8}
\end{eqnarray}
The proper subgraphs as shown in  Figure \ref{two-point-2}
give the contribution (given by the counterterm in eq.
(\ref{one.13.0.4}))
\begin{eqnarray}&& 
{\Gamma}_{J_a^\mu J_b^\nu}^{(2)}\Big |_{\rm SUBGRAPHS}=
\nonumber\\&&
 =
-  i\delta_{ab}
\frac{2}{D-4}\frac{4}{3(4\pi)^2}\frac{g^4}{m_D^4}
\frac{1}{m^2}
\frac{-p^2}{(D-1)}(-p^2 g_{\mu\nu}+p_\mu p_\nu)
\nonumber\\&&
\frac{i}{[-p^2]^{2-\frac{D}{2}}}
\frac{\Gamma\left(2-\frac{D}{2}\right)}{(4\pi)^{\frac{D}{2}}}
\frac{\Gamma\left(\frac{D}{2}-1\right)^2}{\Gamma(D-2)}.
\label{two-loop.11}
\end{eqnarray}
The double poles do not cancel. Then one needs a counterterm
\begin{eqnarray}
{\widehat\Gamma}^{(2)}[JJ]
 =
\frac{1}{(D-4)^2}\left(\frac{m}{m_D}\right)^2
\int d^Dx J_a^\mu
\frac{4g^4}{9(4\pi)^4m^6}\Box(\Box g_{\mu\nu}-\partial_\mu \partial_\nu)
J_a^\nu.
\label{two-loop.12.1}
\end{eqnarray}
A straightforward calculation shows that the sum of all
the contributions in eqs. (\ref{two-loop.8}), (\ref{two-loop.11}) 
and (\ref{two-loop.12.1}) have no double poles. Also the
single pole is zero (this guarantees that sugraph counterterms
do not induce non local terms). Only the finite
part remains (second graph in Figure \ref{two-point-2})
\begin{eqnarray}
&&
\Gamma_{J_a^\mu J_b^\nu}^{(2)}\Big|_{\rm RENORM}
=\delta_{ab}(-p^2 g_{\mu\nu}+p_\mu p_\nu)[-p^2]
\frac{g^4}{m^6(4\pi)^4} 
\nonumber\\&&
\Big\{
\frac{2}{9}\ln^2\Big[\frac{-p^2}{m^2(4\pi)}\Big]
-
\frac{4}{3}\ln\Big[\frac{-p^2}{m^2(4\pi)}\Big]
[-\frac{\gamma}{3}+\frac{2}{9}]
\nonumber\\&&
+ 2 \Big(-\frac{\gamma}{3}+\frac{2}{9}\Big)^2
\Big\}.
\label{two-loop.19p}
\end{eqnarray}
The finite part of the first graph in Figure \ref{two-point-2}
is
\begin{eqnarray}&&
\Gamma_{J_a^\mu J_b^\nu}^{(2)}\Big|_{\rm RENORM}
=\delta_{ab}\frac{1}{3}
 \left(\frac{g^2}{m^3(4\pi)^2}\right)^2 [-p^2]
\Big(
-p^2g_{\mu\mu'}+3p_{\mu}p_{\mu'}
\Big)
\nonumber\\&&
\Big\{
\ln\left(
\frac{[-p^2]}{m^2(4\pi)}
\right)+
\frac{\gamma}{16} - \frac{29}{128}
\Big\}
+ \delta_{ab}
\frac{5}{24} \left(\frac{g^2}{m^3(4\pi)^2}\right)^2 [-p^2]p_{\mu}p_{\mu'}
\nonumber\\
.
\label{two-loop.20p}
\end{eqnarray}
%
\section{Conclusions and outlook}
We have formulated the nonlinear sigma model
in terms of solutions of a functional equation.
The functional equation is obtained by constructing a flat
connection in terms of the sigma field and then
by considering local gauge transformations.
The vertex function at the tree level provides
the Feynman rules in D dimensions. We demonstrated
that these na{\"\i}ve rules yield amplitudes that
satisfy the functional equations in D dimensions.
This fact suggests a simple strategy
for the renormalization of the model in $D=4$
consisting in the recursive subtraction of the poles
in the Laurent expansion.
However some steps of the proof are still at the level
of conjecture. In particular one needs a general proof
of the statement in Section \ref{sec:more} on the
validity of the solution in D dimensions and also of the
same statement
after the subtraction procedure. The last point however 
might follow from the quantum action principle as
formulated in Reference \cite{Breitenlohner:1977hr} for dimensional
renormalization.
\footnote{
We thank D. Maison for this suggestion.}
\par\noindent
With these provisos the final finite
theory depends only on two parameters ($g$ and $m$), but
the amplitudes are constructed with a particular 
subtraction procedure that is in principle not unique. 
\par
We consider the strategy very promising. It could
open a way to attack other non renormalizable theories
as non-abelian gauge theories with a St\"uckelberg
mass term or also the nonlinear sigma model with fermions. 
\section*{Acknowledgments}

We   gratefully
acknowledge the warm hospitality  of the Center for Theoretical
Physics at MIT, Boston, 
where part of this work has been accomplished,
and the partial financial support in the frame of a six-year scientific 
collaboration between MIT and INFN in honor of Bruno Rossi.
For stimulating discussions we are very thankful to R. Jackiw, D. Maison,
P. Menotti, A. Quadri and A.A. Slavnov.


%
%
%
\appendix
\section{One-loop integrals}
\label{app:one-int}
Some useful formulas are given in this appendix
\begin{eqnarray}&&
I_2(p)\equiv
\int \frac{d^D k}{(2\pi)^D}
\frac{1}{k^2(p+k)^2}
\nonumber\\&&
= i\int_0^1 dx \frac{1}{(4\pi)^{\frac{D}{2}}}
\frac{\Gamma\left(2-\frac{D}{2}\right)}{\Gamma\left(2\right)}
\left[-
p^2x(1-x)
\right]^{(\frac{D}{2}-2)}
\nonumber\\&&
=
\frac{i}{[-p^2]^{2-\frac{D}{2}}}
\frac{\Gamma\left(2-\frac{D}{2}\right)}{(4\pi)^{\frac{D}{2}}}
B(\frac{D}{2}-1,\frac{D}{2}-1)
\nonumber\\&&
=
\frac{i}{[-p^2]^{2-\frac{D}{2}}}
\frac{\Gamma\left(2-\frac{D}{2}\right)}{(4\pi)^{\frac{D}{2}}}
\frac{\Gamma\left(\frac{D}{2}-1\right)^2}{\Gamma(D-2)}
.
\label{integral.28}
\end{eqnarray}
\begin{eqnarray}
\int \frac{d^D k}{(2\pi)^D}
\frac{k_\mu}{k^2(p+k)^2}
= -\frac{p_\mu}{2}I_2(p).
\label{integral.32}
\end{eqnarray}
\begin{eqnarray}
\int \frac{d^D k}{(2\pi)^D}
\frac{k_\mu k_\nu}{k^2(p+k)^2}
= \frac{1}{4(D-1)}I_2(p)(-p^2 g_{\mu\nu}+Dp_\mu p_\nu).
\label{integral.35}
\end{eqnarray}
%
\begin{eqnarray}
\langle 0|(\phi_a(x){\phi_b}(y))_+|0\rangle~~
\langle 0|(\phi_c(x){\phi_d}(y))_+|0\rangle
=-\delta_{ab}\delta_{cd}
~ I_2(x-y),
\label{one.13.8}
\end{eqnarray}
where
\begin{eqnarray}
I_2(x)=\int\frac{d^Dp}{(2\pi)^{D}}~ e^{-ipx} I_2(p),
\label{one.13.9}
\end{eqnarray}
and the relations 
\begin{eqnarray} &&
\langle 0|(\partial_\nu\phi_a(x){\phi_b}(y))_+|0\rangle~~
\langle 0|(\phi_c(x){\phi_d}(y))_+|0\rangle
=-\delta_{ab}\delta_{cd}\frac{1}{2}\partial_{\nu}I_2(x-y)
\nonumber\\&&
\langle 0|(\partial_\mu\partial_\nu\phi_a(x){\phi_b}(y))_+|0\rangle~~
\langle 0|(\phi_c(x){\phi_d}(y))_+|0\rangle
\nonumber\\&& 
=\delta_{ab}\delta_{cd}\frac{1}{4(D-1)}(\Box g_{\mu\nu}
-D\partial_\mu \partial_\nu)I_2(x-y)
\nonumber\\&& 
\langle 0|(\partial_\mu\phi_a(x){\phi_b}(y))_+|0\rangle~~
\langle 0|(\partial_\nu\phi_c(x){\phi_d}(y))_+|0\rangle
\nonumber\\&& 
=-\delta_{ab}\delta_{cd}\frac{1}{4(D-1)}(\Box g_{\mu\nu}
+(D-2)\partial_\mu \partial_\nu)I_2(x-y) 
.
\label{one.13.10}
\end{eqnarray}
The pole part of $I_2(x-y)$ is%
\begin{eqnarray}
\left .
I_2(x-y)\right|_{\rm POLE} =
-i\frac{2}{D-4} \frac{1}{(4\pi)^2}\delta_D(x-y).
\label{one.44}
\end{eqnarray}
More  identities
\begin{eqnarray} &&
\langle 0|(\partial_\mu\phi_a(x){\partial^\mu\phi_b}(y))_+|0\rangle~~
\langle 0|(\partial_\nu\phi_c(x){\phi_d}(y))_+|0\rangle =0
\nonumber\\&& 
\langle 0|(\partial_\mu\partial_\nu\phi_a(x){\phi_b}(y))_+|0\rangle~~
\langle 0|(\partial^\mu\phi_c(x){\phi_d}(y))_+|0\rangle 
\nonumber\\&&
=-\delta_{ab}\delta_{cd}\frac{1}{4}\Box \partial_\nu I_2(x-y)
\label{one.42.1}
\end{eqnarray}
and finally
\begin{eqnarray} &&
\langle 0|(\partial_\mu\partial_\nu\phi_a(x){\phi_b}(y))_+|0\rangle~~
\langle 0|(\partial^\mu\partial^\nu\phi_c(x){\phi_d}(y))_+|0\rangle 
\nonumber\\&&
=-\delta_{ab}\delta_{cd}\frac{1}{4}\Box^2  I_2(x-y).
\label{one.43.1}
\end{eqnarray}
%
\section{Two-loop integrals}
\label{app:two-int}
Now we evaluate the two-loop integrals:
\begin{eqnarray}
I_3(p_1) \equiv
\int \frac{d^D p_2}{(2\pi)^D}
\int \frac{d^D p_3}{(2\pi)^D} \frac{1}{p_2^2p_3^2(p_1+p_2+p_3)^2}.
\label{integral.38.0}
\end{eqnarray}
From eq. (\ref{integral.28}) we get
\begin{eqnarray}&& 
I_3(p_1)
=
\frac{i}{(4\pi)^{\frac{D}{2}}}
\frac{\Gamma\left(2-\frac{D}{2}\right)}{\Gamma\left(2\right)}
B(\frac{D}{2}-1,\frac{D}{2}-1)
\nonumber\\&&
\int\frac{d^D p_2}{(2\pi)^{D}} 
\frac{1}{[-(p_1+p_2)^2]^{2-\frac{D}{2}}}
\frac{(-1)}{(-p_2^2)}
\nonumber\\&&
=
\frac{-i}{(4\pi)^{\frac{D}{2}}}
\frac{\Gamma\left(2-\frac{D}{2}\right)}{\Gamma\left(2\right)}
B(\frac{D}{2}-1,\frac{D}{2}-1) 
\nonumber\\&&
\int\frac{d^D p_2}{(2\pi)^{D}}
\int_0^1 dw~ \frac{w^{1-\frac{D}{2}}}
{[-(p_2+wp_1)^2-p_1^2(w-w^2)]^{3-\frac{D}{2}}}
B(2-\frac{D}{2},1)^{-1}
\nonumber\\&&
=
\frac{1}{(4\pi)^{D}}
\frac{\left[\Gamma\left(\frac{D}{2}-1\right)\right]^3
}
{\Gamma(\frac{3D}{2}-3)}
\frac{\Gamma(3-D)}{[-p_1^2]^{3-D}}
\label{integral.38}
\end{eqnarray}
The only pole in D=4 is in $\Gamma(3-D)$.
Now we start from the identity
\begin{eqnarray}
\langle p_{2\mu} \rangle_3
= \langle p_{3\mu} \rangle_3
\label{integral.39}
\end{eqnarray}
and in the integral over $ p_{3}$ we use eq. (\ref{integral.32})
\begin{eqnarray}&&
\langle p_{2\mu} \rangle_3
= - \frac{1}{2}\langle( p_{1\mu}+ p_{2\mu}) \rangle_3
\nonumber\\&&
=- \frac{I_3(p_1)}{2} p_{1\mu}
 - \frac{1}{2}\langle p_{2\mu} \rangle_3
\label{integral.40}
\end{eqnarray}
and then
\begin{eqnarray}
\langle p_{2\mu} \rangle_3
=- \frac{I_3(p_1)}{3} p_{1\mu}.
\label{integral.41}
\end{eqnarray}
In similar fashion we get
\begin{eqnarray}
\langle p_{2\mu} p_{2\nu}\rangle_3
= \langle p_{3\mu}p_{3\nu} \rangle_3
\label{integral.42}
\end{eqnarray}
and in the integral over $ p_{3}$ we use eq. (\ref{integral.35})
\begin{eqnarray}&&
\langle p_{2\mu} p_{2\nu}\rangle_3
= \langle
\frac{1}{4(D-1)}\left(-(p_1+p_2)^2g_{\mu\nu}
+D (p_{1}+ p_2)_\mu(p_{1}+ p_2)_\nu\right)  \rangle_3
\nonumber\\&&
= \frac{1}{4(D-1)}\langle
\left(-p_1^2(1-\frac{2}{3})g_{\mu\nu}
+D p_{1\mu} p_{1\nu}(1-\frac{2}{3})+D p_{2\mu} p_{2\nu}\right)  \rangle_3
\label{integral.43}
\end{eqnarray}
i.e.
\begin{eqnarray}
\langle p_{2\mu} p_{2\nu}\rangle_3
=
\frac{1}{3(3D-4)}\left(-p_1^2g_{\mu\nu}
+D p_{1\mu} p_{1\nu}\right)I_3(p_1).
\label{integral.44}
\end{eqnarray}
\bibliography{reference}

\end{document}